\documentclass[a4paper,twocolumn,11pt,accepted=2021-04-29]{quantumarticle}

\pdfoutput=1
\usepackage[utf8]{inputenc}
\usepackage[english]{babel}
\usepackage[LGR,T1]{fontenc}
\usepackage{amsmath}
\usepackage{float}
\usepackage{graphicx}
\usepackage{amsmath,etoolbox}
\usepackage{amssymb}
\usepackage{xcolor}
\usepackage{hyperref}
\usepackage{soul}
\usepackage{bbm}
\usepackage{commath}
\usepackage{enumerate}
\usepackage{mathtools}
\usepackage{calrsfs}
\DeclareMathAlphabet{\pazocal}{OMS}{zplm}{m}{n}
\usepackage{upgreek}
\usepackage{calc}
\usepackage{accents}

\usepackage[numbers,sort&compress]{natbib}
\graphicspath{{fig/}} 

\newcommand{\declarebsfgreek}[2]{%
  \protected\csdef{bsf#1}{\mathord{\text{\bsfgreekfont#2}}}%
}
\newcommand{\bsfgreekfont}{\usefont{LGR}{cmss}{bx}{it}}
\declarebsfgreek{alpha}{a}
\declarebsfgreek{beta}{b}
\declarebsfgreek{gamma}{g}
\declarebsfgreek{delta}{d}
\declarebsfgreek{epsilon}{e}
\declarebsfgreek{zeta}{z}
\declarebsfgreek{eta}{h}
\declarebsfgreek{theta}{j}
\declarebsfgreek{iota}{i}
\declarebsfgreek{kappa}{k}
\declarebsfgreek{lambda}{l}
\declarebsfgreek{mu}{m}
\declarebsfgreek{nu}{n}
\declarebsfgreek{xi}{x}
\declarebsfgreek{omicron}{o}
\declarebsfgreek{pi}{p}
\declarebsfgreek{rho}{r}
\declarebsfgreek{sigma}{s}
\declarebsfgreek{tau}{t}
\declarebsfgreek{upsilon}{u}
\declarebsfgreek{phi}{f}
\declarebsfgreek{chi}{q}
\declarebsfgreek{psi}{u}
\declarebsfgreek{omega}{w}

\DeclareMathAlphabet{\pazocal}{OMS}{zplm}{m}{n}

\DeclareMathOperator{\tr}{tr}

\newcommand{\comments}[1]{}
\newcommand{\bea}{\begin{eqnarray}}
\newcommand{\eea}{\end{eqnarray}}

\newcommand{\rvline}{\hspace*{-\arraycolsep}\vline\hspace*{-\arraycolsep}}
\newcommand{\iu}{{i\mkern1mu}}

\makeatletter
\DeclareFontFamily{OMX}{MnSymbolE}{}
\DeclareSymbolFont{MnLargeSymbols}{OMX}{MnSymbolE}{m}{n}
\SetSymbolFont{MnLargeSymbols}{bold}{OMX}{MnSymbolE}{b}{n}
\DeclareFontShape{OMX}{MnSymbolE}{m}{n}{
	<-6>  MnSymbolE5
	<6-7>  MnSymbolE6
	<7-8>  MnSymbolE7
	<8-9>  MnSymbolE8
	<9-10> MnSymbolE9
	<10-12> MnSymbolE10
	<12->   MnSymbolE12
}{}
\DeclareFontShape{OMX}{MnSymbolE}{b}{n}{
	<-6>  MnSymbolE-Bold5
	<6-7>  MnSymbolE-Bold6
	<7-8>  MnSymbolE-Bold7
	<8-9>  MnSymbolE-Bold8
	<9-10> MnSymbolE-Bold9
	<10-12> MnSymbolE-Bold10
	<12->   MnSymbolE-Bold12
}{}

\def\D{\mathrm{d}}

\newcommand{\ignore}[1]{}
\newcommand{\nobibentry}[1]{{\let\nocite\ignore\bibentry{#1}}}
\newcommand{\re}[1]{\text{Re}\,#1}

\newcommand{\ket}[1]{\left\vert#1\right\rangle}
\newcommand{\bra}[1]{\left\langle#1\right\vert}

\begin{document}

\title{Local master equations bypass the secular approximation}

\author{Stefano Scali}
\affiliation{Department of Physics and Astronomy, University of Exeter, Exeter EX4 4QL, United Kingdom}
\email{s.scali@exeter.ac.uk}

\author{Janet Anders}
\affiliation{Department of Physics and Astronomy, University of Exeter, Exeter EX4 4QL, United Kingdom}
\affiliation{Institut f\"ur Physik und Astronomie, University of Potsdam, 14476 Potsdam, Germany.}
\email{janet@qipc.org}

\author{Luis A. Correa}
\affiliation{Department of Physics and Astronomy, University of Exeter, Exeter EX4 4QL, United Kingdom}
\email{l.correa-marichal@exeter.ac.uk}
 
\begin{abstract}
Master equations are a vital tool to model heat flow through nanoscale thermodynamic systems. Most practical devices are made up of interacting sub-systems and are often modelled using either \emph{local} master equations (LMEs) or \emph{global} master equations (GMEs). While the limiting cases in which either the LME or the GME breaks down are well understood, there exists a `grey area' in which both equations capture steady-state heat currents reliably but predict very different \emph{transient} heat flows. In such cases, which one should we trust? Here we show that, when it comes to dynamics, the local approach can be more reliable than the global one for weakly interacting open quantum systems. This is due to the fact that the \emph{secular approximation}, which underpins the GME, can destroy key dynamical features. To illustrate this, we consider a minimal transport setup and show that its LME displays \emph{exceptional points} (EPs). These singularities have been observed in a superconducting-circuit realisation of the model \cite{partanen2019exceptional}. However, in stark contrast to experimental evidence, no EPs appear within the global approach. We then show that the EPs are a feature built into the Redfield equation, which is more accurate than the LME and the GME. Finally, we show that the local approach emerges as the weak-interaction limit of the Redfield equation, and that it entirely avoids the secular approximation.
\end{abstract}

\maketitle


\section{Introduction}

Master equations and quantum thermodynamics go hand in hand. The former have become essential tools to make sense of the `thermodynamics' of quantum systems. But, conversely, the early works on quantum thermodynamics \cite{alicki1979engine,spohn1978entropy,kosloff1984quantum} focused on the study of the mathematical properties of master equations. Nowadays the field is evolving very rapidly \cite{binder2018thermodynamics}, and quantum heat devices are making the transition from theory to experiments on a wide range of platforms, including trapped ions, solid-state systems, atomic gases, single-electron systems, nanoscale thermoelectrics, and superconducting circuits \cite{rossnagel2016single,von2019spin,klatzow2019experimental,gelbwaser2015laser,zou2017quantum,koski2014experimental,masuyama2018information,naghiloo2018information,cottet2017observing}. 

The Gorini--Kossakowski--Sudarshan--Lindblad (GKSL) quantum master equation \cite{gorini1976completely,lindblad1976generators} makes it easy to draw parallels between the dissipative dynamics of a single open quantum system and the thermodynamics of macroscopic devices \cite{alicki2019introduction,Kosloff2013,Kosloff2014}. Namely, these equations can be derived from first principles in the limit of weak system--environment coupling, and may lead to \emph{thermal equilibrium} \cite{spohn1977algebraic}. We shall refer to such `thermalising' GKSL equations as \emph{global master equations} (GMEs). Furthermore, heat currents can be formally defined such that they obey the second law of (classical) thermodynamics \cite{alicki1979engine,alicki2019introduction}. However, the underlying assumptions of the global master equation require a clean timescale separation \cite{breuer2002theory}, which may break down for, e.g., small multipartite quantum-thermodynamic devices that interact \emph{weakly} among them (see, e.g.,\cite{gonzalez2017testing,Hofer2017}) and large many-body open quantum systems \cite{PhysRevE.76.031115}.

\begin{figure}
    \centering
    \includegraphics[width=0.9\linewidth]{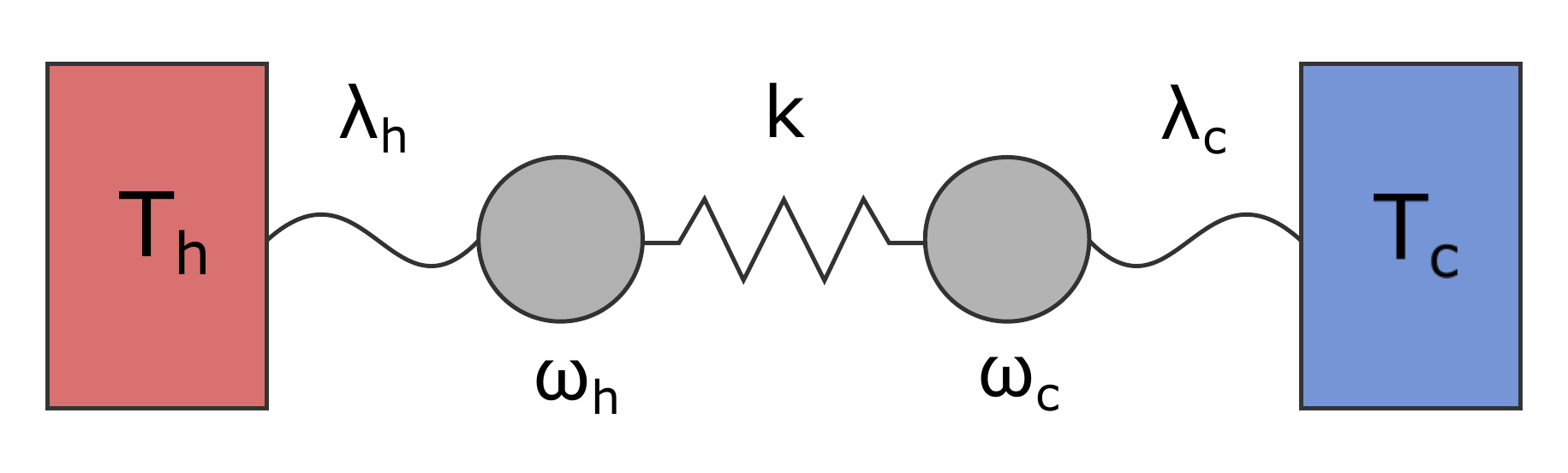}
    \includegraphics[width=0.9\linewidth]{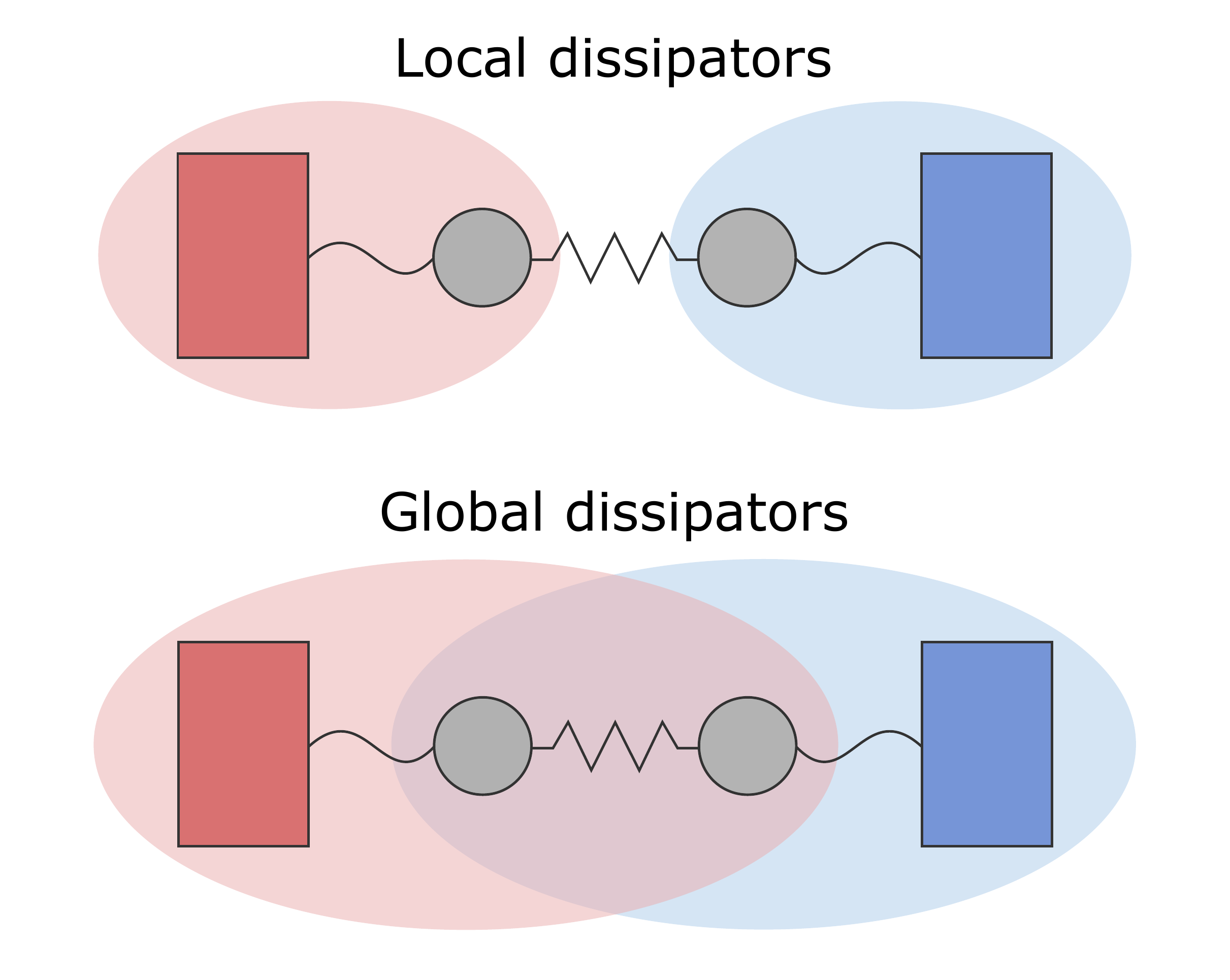}
    \caption{
    \textbf{Example system and local/global pictures.} Two resonators of frequencies $\omega_h$ and $\omega_c$ are coupled linearly with strength $k$, and connected weakly with dissipation strengths $\lambda_h$ and $\lambda_c$ to two independent thermal reservoirs at temperatures, $T_h$ and $T_c$, respectively \textbf{(top)}. Heat thus flows through the resonators. The key difference between the local \textbf{(middle)} and the global \textbf{(bottom)} master equations is illustrated by the shaded regions that indicate the scope of the hot and cold dissipators, respectively.}
    \label{fig:setup_full}
\end{figure}

Alternatively, in multipartite open quantum systems, master equations have often been built heuristically by `adding up' GKSL terms (cf. Fig.~\ref{fig:setup_full}). These are referred-to as \emph{local master equations} (LMEs). While such equations do comply, by construction, with the minimum expectation of generating a completely positive dynamics, they have been criticised for their thermodynamic deficiencies  \cite{rivas2010markovian,correa2013performance,levy2014local,manrique2015nonequilibrium,stockburger2016thermodynamic,PhysRevE.94.062143,gonzalez2017testing,mitchison2018non,kolodynski2018adding,naseem2018thermodynamic,cattaneo2019local,mcconnell2019electron,hewgill2020quantum}. Namely, unlike global master equations, LMEs fail to bring systems to thermal equilibrium, even in the limit of weak system--environment interactions \cite{stockburger2016thermodynamic,kolodynski2018adding}. When applied to quantum heat devices, they entirely miss crucial physics, such as heat leaks and internal dissipation \cite{correa2013performance,correa2015internal}. They may even predict flagrant violations of the Second Law of thermodynamics, in the form of cold-to-hot stationary heat flows \cite{levy2014local}. Surprisingly, however, the LME does prove very accurate in some cases---even more accurate than the GME \cite{gonzalez2017testing,Hofer2017}. The aim of this paper is to understand \emph{when} and \emph{why}. Using the most suitable master equation in each situation (see Fig.~\ref{fig:setup_full}) can make a crucial difference when studying the thermodynamics of any nanoscale heat device. 

Local master equations may be understood as a rough approximation to the true dissipative dynamics, valid in the limit of weak interactions between the sub-systems. Starting from a microscopic model this may be shown in two closely related ways---either by carefully introducing a coarse-graining in the time-evolution of the open system \cite{schaller2008preservation,cresser2017coarse,seah2018refrigeration,farina2019recovering,farina2020going,majenz2013coarse,lidar2001completely,mozgunov2020completely,elouard2020thermodynamics}, or by truncating a perturbative expansion of the master equation in the internal coupling strength \cite{trushechkin2016perturbative,purkayastha2016out}.  

The disagreement between the steady-state thermodynamic predictions of GME and LME had been illustrated before \cite{manrique2015nonequilibrium,correa2012asymptotic,Hofer2017}. Here, we put the spotlight on situations in which they agree in the steady state, but differ during the transient dynamics. In such cases, which master equation is correct? One that is thermodynamically sound (GME), or a truncated series expansion with limited validity and serious thermodynamic deficiencies (LME)? Strikingly, our answer is that \emph{we should always trust the latter} within its error bars, which assume weak internal couplings. This holds for any multipartite quantum heat device which uses \emph{frequency filters} to couple to the environment \cite{Correa2014}, e.g., a qubit or a harmonic oscillator.

For illustration, we focus on the specific model of two coupled resonators that connect two thermal reservoirs. We find that the LME exhibits a family of \emph{exceptional points} \cite{kato2013perturbation} (EPs) in its dynamics, while the corresponding GME does not. Yet, EPs have indeed been found experimentally in a superconducting circuit described by the same model \cite{partanen2019exceptional}. In addition, at weak internal couplings, the transient heat currents obtained from the LME agree with the much more accurate Redfield equation \cite{redfield1957theory} while \emph{not} with the global approach. The reason for the failures of the GME is that it is underpinned by the (somewhat crude) secular approximation, which misses relevant physics. In contrast, the LME can be obtained directly from the Redfield equation, bypassing the secular approximation. Our results thus add much needed clarity to the long-standing `local-versus-global' debate, and explain various previously reported features of both the global and the local approach.

This paper is structured as follows: We begin by giving an overview of open-system dynamics within the global and local approach in Sec.~\ref{sec:open-system_dynamics} and then, describing the details of the example model in Sec.~\ref{sec:the_system}. In Sec.~\ref{sec:EPs} we introduce the concept of \emph{exceptional points} and discuss how to search for them, given the equations of motion of a linear open quantum system. We then illustrate the different exceptional-point structure in parameter space according to the local, global, and Redfield approaches (Sec.~\ref{sec:results-and-discussion}). Finally, in Sec.~\ref{sec:heat-currents}, we show that, in resonance, the local approach succeeds at capturing the correct heat-flow dynamics, while the global master equation fails. In Sec.~\ref{sec:conclusions} we summarise and conclude.


\section{Open-system dynamics}\label{sec:open-system_dynamics}

\subsection{The global master equation}\label{sec:global_master_equation}

The Hamiltonian of a \emph{generic} multipartite open system connected to various independent bosonic heat baths reads
\begin{equation}
    \pmb H = \pmb H_S + \pmb H_B + \pmb H_{SB},
\end{equation}
where $\pmb H_S$ is the Hamiltonian of the system, $\pmb H_B$ the Hamiltonian of the baths, each made of infinitely many uncoupled harmonic oscillators, and $\pmb H_{SB}$ stands for the interaction between system and baths. Specifically, let 
\begin{equation}
    \pmb H_S = \sum\nolimits_i \pmb H^\text{(loc)}_i + k\,\pmb V,
\end{equation}
where $ \pmb H_i^\text{(loc)} $ is the local Hamiltonian of each sub-system and $ \pmb V $ denotes the interactions between them. The parameter $ k $ controls the magnitude of the latter. Every sub-system couples to its own independent bath, i.e.
\begin{equation}\label{eq:H_SB}
    \pmb H_{SB} = \sum\nolimits_\alpha \lambda_\alpha\,\pmb S_\alpha\otimes \pmb B_\alpha,
\end{equation}
where $ \pmb B_\alpha $ is a generic bath operator and  $ \pmb S_\alpha $ is a system operator which does not commute with $\pmb H_S$, thus allowing for energy \emph{dissipation} as well as \emph{decoherence}. Here, $\lambda_\alpha$ controls the strength of the interaction with bath $\alpha$.

The effective equation of motion for any arbitrary system observable $ \pmb O $ can be cast in the standard GKSL form \cite{gorini1976completely,lindblad1976generators}.
Although its microscopic derivation is textbook material \cite{breuer2002theory}, we provide it in Appendix~\ref{app:clean_derivation_master_equation}, making as few assumptions as possible. Essentially, these are:
\begin{enumerate}[(i)]
    \item that the dissipation strengths are small,
    \item that system and bath start uncorrelated,
    \item that the bath correlation functions are short-lived,
    \item and that there is a clear-cut timescale separation between (fast) coherent and (slow) dissipative processes.
\end{enumerate}

The so-called (partial) Redfield equation \cite{redfield1957theory} (cf. Appendix~\ref{app:clean_derivation_master_equation}) follows from these assumptions,
\begin{equation}\label{eq:redfield_main_text}
    \frac{\D\pmb O}{\D t} = i[\pmb H_S,\pmb O] + \sum\nolimits_\alpha\mathcal{R}^\dagger_\alpha(\pmb O) + \pazocal{O}(\lambda_\alpha^3).
\end{equation}
The super-operator $\mathcal{R}^\dagger_\alpha(\pmb O)$ associated to bath $\alpha$ is given by
\begin{multline}\label{eq:redfield_main_text_2}
    \mathcal{R}_\alpha^\dagger(\pmb O) = \frac12\sum_{\omega_i\times\omega_j < 0}\gamma_{\omega_i}^{(\alpha)}\Big(\pmb{A}_{\omega_i}^{(\alpha)\dagger}\pmb{O}\pmb{A}_{\omega_j}^{(\alpha)} \\
    -\pmb{A}_{\omega_j}^{(\alpha)}\pmb{A}_{\omega_i}^{(\alpha)\dagger}\pmb{O}\Big) + \text{h.c.}
\end{multline}
where the summation $\omega_i\times\omega_j < 0$ runs over the frequencies $\omega_i$ and $\omega_j$ of all \emph{open decay channels} \cite{breuer2002theory} with different sign. The decay rates $ \gamma_\omega^{(\alpha)} $ are
\begin{multline}\label{eq:decay_rate}
    \gamma_\omega^{(\alpha)} 
    = 2\,\lambda_\alpha^2 \, \re\int_0^{\infty}\D s\,e^{i\omega s} \, \langle \pmb B_\alpha(s)\pmb B_\alpha(0) \rangle ,
\end{multline}
where the notation $\langle \cdot \rangle$ indicates thermal averaging at temperature $T_\alpha$. Finally, $\pmb{A}_\omega^{(\alpha)}$ is the `jump' operator of the decay channel at frequency $\omega$ from bath $\alpha$. These satisfy
\begin{subequations}\label{eq:jump_operators_properties}
    \begin{align}
        [\pmb H_S,\pmb A_\omega^{(\alpha)}] &= -\omega\,\pmb A_\omega^{(\alpha)}, \label{eq:jump_operator_commutation}\\
        \pmb S_\alpha &=  \sum\nolimits_\omega \pmb A_\omega^{(\alpha)}, \\
        \pmb A_\omega^{(\alpha)\,\dagger} &= \pmb{A}_{-\omega}^{(\alpha)}.
    \end{align}
\end{subequations}
In particular, Eq.~\eqref{eq:decay_rate} implies that $ \mathcal{R}^\dagger_\alpha $ is $\pazocal{O}(\lambda_\alpha^2)$. Note as well that, to make sense of \eqref{eq:redfield_main_text} and all the following adjoint master equations, we must always take expectation values, as we do in Eqs.~\eqref{eq:first-order-lme}, \eqref{eq:coeffs_gme-original-basis-eqs} and \eqref{eq:coeffs-eqs-redfield}. 

Pushing (iv) to its last consequences justifies the (full) \emph{secular approximation}, and allows to bring Eq.~\eqref{eq:redfield_main_text} to the simpler GKSL form,
\begin{equation}\label{eq:GKSL_general}
    \frac{\D\pmb O}{\D t} = i[\pmb H_S,\pmb O] + \sum\nolimits_\alpha\mathcal{G}^\dagger_\alpha(\pmb O) + \pazocal{O}(\lambda_\alpha^3), 
\end{equation}
where the super-operators $ \mathcal{G}^\dagger_\alpha(\pmb O) $ are given by
\begin{multline}\label{eq:global_dissipator}
    \mathcal{G}_\alpha^\dagger(\pmb O) = \sum\nolimits_\omega \gamma_\omega^{(\alpha)}\left(\pmb A_\omega^{(\alpha)\,\dagger} \pmb O \pmb A_\omega^{(\alpha)}\right. \\ \left. - \frac12 \lbrace \pmb A_\omega^{(\alpha)\,\dagger} \pmb A_\omega^{(\alpha)},\pmb O \rbrace_+\right) .  
\end{multline}
Here, $ \{\cdot,\cdot\}_+ $ stands for the anti-commutator. Eq.~\eqref{eq:GKSL_general} is a \emph{global} master equation. The name tag highlights the fact that the $ \pmb A_\omega^{(\alpha)} $ enable jumps between eigenstates of the full multipartite $ \pmb H_S $, rather than between states of each sub-system $ \pmb H^\text{(loc)}_\alpha $ (see Fig.~\ref{fig:setup_full}).

Since full diagonalisation of $ \pmb H_S $ is needed to construct these jump operators, setting up Eq.~\eqref{eq:GKSL_general} may become computationally unworkable; especially, when one wishes to scale up a many-body open quantum system. Furthermore, the GME suffers from another important issue, especially when applied to systems with a \emph{dense} energy spectrum. In such cases, assumption (iv) from the list above is likely to break down \cite{PhysRevE.76.031115}, which could invalidate the GME's predictions \cite{gonzalez2017testing,Hofer2017,strasberg2018fermionic}. On the `plus side', constructing jump operators that fulfil Eqs.~\eqref{eq:jump_operators_properties} is guaranteed to bring the system into a state of thermodynamic equilibrium whenever all temperatures coincide, i.e., $ T_\alpha = T~\forall \alpha$. This is analogous to the Zeroth Law of thermodynamics \cite{planck1901treatise}. Since, in addition, the dynamical map resulting from \eqref{eq:GKSL_general} is \emph{completely positive} \cite{spohn1978entropy}, it can be shown that  
\begin{subequations}\label{eq:Clausius}
    \begin{align}
        &\sum\nolimits_\alpha \frac{\dot{\pazocal{Q}}_\alpha}{T_\alpha} \leq 0,\label{eq:clausius-ineq}\\
        \mbox{for}~~ &\dot{\pazocal{Q}}_\alpha \coloneqq \langle\mathcal{G}^\dagger_\alpha(\pmb H_S)\rangle_\infty, \label{eq:heat-current-definition}
    \end{align}
\end{subequations}
where the steady-state heat currents $\dot{\pazocal{Q}}_\alpha$ account for the stationary rate of energy influx from bath $ \alpha $ (i.e., $\langle \cdot \rangle_\infty$ denotes here \emph{stationary} average). Eq.~\eqref{eq:Clausius} is interpreted as the Second Law of thermodynamics, since it is formally identical to the statement of Clausius' theorem \cite{alicki2019introduction,alicki1979engine}. Therefore, the global GKSL equation is particularly well suited to study the thermodynamics of (weakly dissipative) open quantum systems.

\subsection{The local master equation} \label{sec:local_master_equations}

Another type of quantum master equation may be built by simply \emph{adding up} dissipators that act locally on a specific part of the system
\begin{equation}\label{eq:LME}
    \frac{\D\pmb O}{\D t} = i[\pmb H_S,\pmb O] + \sum\nolimits_\alpha\mathcal{L}_\alpha^\dagger(\pmb O),
\end{equation}
where $\mathcal{L}_\alpha^\dagger(\pmb O)$ takes the form
\begin{multline}\label{eq:local_dissipator}
    \mathcal{L}_\alpha^\dagger(\pmb O) = \sum\nolimits_{\omega'} \gamma_{\omega'}^{(\alpha)}\left( \pmb L_{\omega'}^{(\alpha)\,\dagger}\pmb O \pmb L_{\omega'}^{(\alpha)} \right. \\  - \left.\frac12\{ \pmb L^{(\alpha)\,\dagger}_{\omega'} \pmb L^{(\alpha)}_{\omega'},\pmb O \}_+ \right).
\end{multline}
Here, the non-hermitian operators $\pmb L_{\omega'}^{(\alpha)}$ have the properties,
\begin{subequations}
    \begin{align}
        [\pmb H_\alpha^\text{(loc)},\pmb L_{\omega'}^{(\alpha)}] &= -\omega'\pmb L_{\omega'}^{(\alpha)},
        \label{eq:local_dissipators_commutation}\\
        \pmb S_\alpha &= \sum\nolimits_{\omega'} \pmb L_{\omega'}^{(\alpha)}, \\
        \pmb L_{\omega'}^{(\alpha)\,\dagger} &= \pmb{L}_{-\omega'}^{(\alpha)}.
    \end{align}
\end{subequations}
in place of Eqs.~\eqref{eq:jump_operators_properties}. Note that the sum does not run over the Bohr frequencies $ \omega $ of $ \pmb H_S $, but over those of $ \sum\nolimits_\alpha\pmb H_\alpha^\text{(loc)} $; hence the notation $ \omega' $. 

Such local equation is easily \emph{scalable}, as the resonators must be diagonalised individually when searching for the operators $ \pmb L_{\omega'}^{(\alpha)} $. It also shares with the global master equation its GKSL form, which means that the resulting dynamics is, again, completely positive. 

`Completely positive' is often equated identically to `physical' because at least it produces positive probabilities. We remark, however, that (completely positive) LMEs violate the Zeroth Law by construction \cite{stockburger2016thermodynamic}. Indeed, due to Eq.~\eqref{eq:local_dissipators_commutation}, the dissipators $ \mathcal{L}_{\alpha}^\dagger $ `try' to pull the system towards the local thermal state $ \propto \exp{(- \sum_\alpha\,\pmb H^{\text{(loc)}}_\alpha/T_\alpha)} $, which does not commute with the full $ \pmb H_S $ appearing in the `coherent-evolution' term of Eq.~\eqref{eq:LME}. Hence, according to the LME, the system would never thermalise, even if all temperatures $T_\alpha$ are identical. As a result, also the Second Law as stated in Eq.~\eqref{eq:Clausius}, may be violated. In fact, in our example model of Eq.~\eqref{eq:H_non_RWA}, the local approach invariably predicts unphysical cold-to-hot heat currents whenever $ \omega_c/T_c < \omega_h/T_h $ \cite{levy2014local}. This, however, would cease to be an issue if the environments were \textit{active}, i.e., able to exchange both heat and work with the system \cite{de2018reconciliation}. 

\subsection{Comparing local and global approach} \label{sec:localvsglobal}

A local master equation can be motivated physically beyond merely ``adding up dissipators''. For instance, we may obtain the LME directly from a formal collisional model, in the limit of instantaneous collisions \cite{scarani2002thermalizing,barra2015thermodynamic}. However, in this case, the thermodynamics of the LME needs to be reassessed to account for the `cost' of those collisions with bath `units' \cite{barra2018smallest,de2018reconciliation}. On the other hand, for microscopic Hamiltonian models, the general Markovian weak-coupling master equation can be simplified by \emph{coarse-graining} over a relevant time-scale \cite{schaller2008preservation,cresser2017coarse,seah2018refrigeration,farina2019recovering,farina2020going,majenz2013coarse,lidar2001completely,mozgunov2020completely,elouard2020thermodynamics}. Depending on how this averaging is done, both the LME and the GME can emerge naturally \cite{cresser2017coarse,seah2018refrigeration}.  

The LME \eqref{eq:LME} can be alternatively viewed as the lowest order in an expansion of the GME \eqref{eq:GKSL_general} \cite{trushechkin2016perturbative,purkayastha2016out}. Namely, applying perturbation theory to the eigenstates and eigenvalues of $ \pmb H_S $ for small $ k $ and following the procedure to obtain a GME \cite{breuer2002theory,trushechkin2016perturbative}, gives
\begin{equation}
    \mathcal{G}^\dagger_\alpha = \mathcal{G}_\alpha^{\dagger\,(0)} + k\,\mathcal{G}_\alpha^{\dagger\,(1)} + k^2\,\mathcal{G}_\alpha^{\dagger\,(2)} + \cdots.
\end{equation}
Whenever $ \mathcal{G}^{\dagger\,(0)}_\alpha = \mathcal{L}_\alpha^\dagger $, one can claim that the LME becomes equivalent to truncating the expansion at zeroth order in $ k $. Since all terms $ \mathcal{G}^{\dagger\,(n)} $  are $ \pazocal{O}(\lambda^2) $, the LME would thus be accurate within $k\lambda^2$-sized `error bars'. If $ k $ is weak enough, this may be acceptable when compared with the intrinsic error of the GME, set at $ \pazocal{O}(\lambda^3) $ (cf. Eq.~\eqref{eq:GKSL_general}). In light of this interpretation, the unphysical cold-to-hot heat flows predicted by the local approach may be thought-of as a mere artifact of the truncated expansion \cite{trushechkin2016perturbative}. 

However, if the degeneracy of $ \pmb H_S $ changes depending on whether $ k = 0 $ or $ k \neq 0 $, the zeroth order of the global dissipator does not coincide with the local one (i.e., $ \mathcal{G}^{\dagger\,(0)}_\alpha \neq \mathcal{L}^\dagger_\alpha $) \cite{trushechkin2016perturbative}. As we will see, this is precisely what happens to our example model \eqref{eq:H_non_RWA} when $ \omega_c = \omega_h = \omega $ \cite{gonzalez2017testing,Hofer2017,cresser2017coarse,benatti2020bath}. Instead, we find that it is the zeroth order of the {\it Redfield} dissipator that coincides with the local one (i.e., $ \mathcal{R}^{\dagger\,(0)}_\alpha = \mathcal{L}^\dagger_\alpha $) \cite{cresser2017coarse}.


\section{The system}\label{sec:the_system}

The system considered in what follows is sketched in Fig.~\ref{fig:setup_full}. It is comprised of two coupled resonators with frequencies $\omega_c$ and $\omega_h$, which we model as
\begin{equation}\label{eq:H_non_RWA}
    \pmb H_S = \sum\nolimits_{\alpha\in\{c,h\}} \left(\frac12\omega_\alpha^2 \pmb x_\alpha^2+ \frac{\pmb p_\alpha^2}{2}\right)  + k\,\pmb x_c\,\pmb x_h
\end{equation}
where $ \pmb x_\alpha $ and $ \pmb p_\alpha $ are the corresponding quadratures, and $ k $ denotes the inter-resonator coupling strength. We also set the masses to $ m_\alpha = 1 $. From now on, we work in units of $\hbar = k_B = 1$. This model can describe the two capacitively coupled superconducting resonators studied experimentally in Ref.~\cite{partanen2019exceptional} in search for exceptional points\footnote{More precisely, the Hamiltonian considered in the analysis of the experiment, i.e.,
\begin{equation*}
\pmb H_S' = \omega_c\,\pmb a_c^\dagger \pmb a_c + \omega_h\,\pmb a_h^\dagger \pmb a_h + \kappa\,(\pmb a_c^\dagger\pmb a_h + \pmb a_c \pmb a_h^\dagger ),    
\end{equation*}
follows directly from our Eq.~\eqref{eq:H_non_RWA} after performing the \textit{rotating-wave approximation} (also referred-to as pre-tracing secular approximation \cite{fleming2010rotating}), discarding the zero-point energy terms, and defining $ k \coloneqq 2\kappa\sqrt{\omega_c\,\omega_h} $. Since we focus on the resonant case $ \omega_h = \omega_c $, using either $ \pmb H_S $ or $ \pmb H_S' $ does not make any difference, as we have verified.}. 

Each resonator is weakly connected to a local heat bath $\alpha\in\{c,h\}$ (`c' for cold and `h' for hot) with equilibrium temperatures $ T_c < T_h $. The resonator--bath couplings are
\begin{align}
    \label{eq:h_int}
    \pmb H_{SB} = \lambda_c\,\pmb x_c \otimes \pmb B_c + \lambda_h\,\pmb x_h \otimes \pmb B_h,
\end{align}
where the bath operators $ \pmb B_\alpha$ are
\begin{align}\label{eq:bath-operator}
    \pmb B_\alpha = \sum\nolimits_\nu g_\nu^{(\alpha)} \boldsymbol{\mathsf{q}}_\nu^{(\alpha)},
\end{align}
and $\boldsymbol{\mathsf{q}}_\nu^{(\alpha)}$ stands for the coordinate of the environmental mode of bath $ \alpha $ at frequency $ \omega_\nu $. Note that $\pmb H_{SB}$ is linear in the dissipation strengths $\lambda_h$ and $\lambda_c$. The couplings $ g_\nu^{(\alpha)} $ can be collected into the \emph{spectral densities}
\begin{equation}
    J_\alpha(\omega) \coloneqq \pi\,\lambda_\alpha^2\sum\nolimits_\nu \frac{g_\nu^{(\alpha)\,2}}{2\omega_\nu}\,\delta(\omega-\omega_\nu),
\end{equation}
where $ \delta(\omega-\omega_\nu) $ stand for Dirac deltas. In the following we will work with the standard Ohmic--algebraic spectral densities
\begin{equation}\label{eq:ohmic-algebraic}
    J_\alpha(\omega) = \lambda^2_\alpha\,\omega\frac{\Lambda^2}{\omega^2+\Lambda^2}
\end{equation}
with $\alpha\in\{c,h\}$. Here, the phenomenological parameter $\Lambda$ sets an `ultraviolet' cut-off on the spectrum of the bath. We note that the precise analytical form of the spectral density does not play an active role in our problem, as long as $ \Lambda $ is large when compared with all other relevant energy scales. Given the spectral densities, the decay rates $ \gamma_\omega^{(\alpha)} $ of Eq.~\eqref{eq:decay_rate} take the form,
\begin{equation}
    \label{eq:decay_rate_system}
    \gamma_\omega^{(\alpha)} = 2 J_\alpha(\omega) \left(1 + \frac{1}{e^{\,\omega/T_\alpha}-1}\right) .
\end{equation}


\section{Exceptional points}\label{sec:EPs}

We now formally introduce the concept of exceptional points (EPs), before drawing the link to open-system dynamics. EPs are branch-point singularities which appear under variation of parameters of non-Hermitian matrices, such as those describing the dynamics of quantum dissipative systems. EPs have been used as resource for applications such as sensitivity amplification in micro-resonators \cite{wiersig2014enhancing,chen2017exceptional}, laser-mode selectivity \cite{hodaei2014parity}, and topological chirality \cite{doppler2016dynamically,xu2016topological}. More recently, a gain in signal-to-noise-ratio has been shown in EP sensors \cite{zhang2019quantum} highlighting the practical relevance of exceptional points.

\subsection{Formal definition and witnesses}\label{sec:EPs-definition-witnesses}

Let $  \mathsf{M}(k) \in \mathbb{C}^{N \times N}$ be an $ N \times N $ matrix dependent on some parameter (or set of parameters) $ k $. We denote the right eigenvectors of this matrix by $ \{ \ket{v_j} \}_{j=1,\cdots,N} $; i.e.,
\begin{equation}
     \mathsf{M}(k)\ket{v_j} = \mu_j \ket{v_j}.
\end{equation}
The corresponding \emph{left} eigenvectors $ \{ (v_j \vert \}_{j=1,\cdots,N} $ are defined instead by
\begin{equation}
     \mathsf{M}(k)^{\mathsf{T}} (v_j\vert^\dagger= \mu_j (v_j\vert^\dagger.
\end{equation}
Notice that $ (v_j\vert^\dagger $ is a \emph{column} vector (like $\ket{v_j}$) due to the Hermitian conjugation. These two indexed families of vectors form a \emph{bi-orthogonal} set \cite{okolowicz2003dynamics}; that is,
\begin{align}
    ( v_i \vert v_j \rangle &= 0 \quad \text{for} ~ i \neq j, \\ \notag
    ( v_i \vert v_i \rangle &\leq 1,
\end{align}
with $ ( v_i \vert v_i \rangle = 1 $ being fulfilled only if $  \mathsf{M}(k) $ is Hermitian. Note that, in general, $( v_i \vert v_i \rangle $ can be negative. We say that the matrix $  \mathsf{M}(k) $ has an exceptional point for those parameter choices $ k $ resulting in $ (v_i \vert v_i \rangle = 0 $ for two or more of the indices $ i \in \{1,\cdots,N\} $. This phenomenon is called \emph{self-orthogonality} and it is the hallmark of the coalescence of two or more right eigenvectors \cite{moiseyev2011non}.

In order to locate the exceptional points of $  \mathsf{M}(k) $ we could search for zeros of the \emph{phase rigidities} $ \upphi_i(k) \coloneqq \vert (v_i|v_i\rangle\vert $ for $ i\in\{1,\cdots,N\} $ as a function of $k$. However, analysing the behaviour of every single eigenvector can be time-consuming for large $ N $. Luckily, we shall only be interested in finding \emph{where} in parameter space an EP of $  \mathsf{M}(k) $ is located, and not in \emph{which} or \emph{how many} eigenvectors coalesce. We can thus exploit the fact that, at an EP, the set $ \{ \ket{v_i} \}_{i=1,\cdots,N} $ does not form a complete basis. Therefore, the matrix $ \mathsf{V}_k = ( \ket{v_1},\cdots,\ket{v_N} )$ features a singularity and the norm of its inverse diverges. Consequently, the \emph{condition number} of $ \mathsf{V}_k $, denoted $\upkappa(\mathsf{V}_k)$, could be a suitable witness for an EP. Namely,
\begin{align}\label{eq:condition-number}
    \upkappa(\mathsf{V}_k) \coloneqq \norm{\mathsf{V}_k}\lVert\mathsf{V}_k^{-1}\rVert,  
\end{align}
where the operator norm $ \norm{\cdot} $ is defined as
\begin{equation}
    \norm{\mathsf{O}} \coloneqq \max\nolimits_{\,\mathsf{x}}{\frac{\norm{\mathsf{O}\,\mathsf{x}}_p}{\norm{\mathsf{x}}_p}}, 
\end{equation}
$ \mathsf{x} $ is an arbitrary non-zero vector, and $ \norm{\cdot}_p $ stands for the $p$-norm
\begin{equation}
    \norm{\mathsf{x}}_p \coloneqq \big(\sum\nolimits_i \vert\mathsf{x}_i\vert^p\big)^{1/p}.
\end{equation}
Here, the parameter $ p $ could take on any real value $ p \geq 1 $. In our numerical calculations, we use the 2-norm. Regardless of $ p $, $ \upkappa(\mathsf{V}_k) $ diverges \emph{iff} there is an EP at position $ k $ in parameter space. Also, note that $ \lVert\mathsf{V}_k^{-1}\rVert $ can be alternatively cast as $ \lVert\mathsf{V}_k^{-1}\rVert = \max_\mathsf{x} \lVert \mathsf{x} \rVert/\lVert \mathsf{V}_k\,\mathsf{x} \rVert $, which resolves the issue of inverting a singular matrix.

\subsection{EPs in open quantum systems}\label{sec:EPs-open-systems}

In many cases of practical interest, an open system may be fully described by choosing a set of observables $ \bsftheta = (\boldsymbol{\upvartheta}_1, \cdots,\boldsymbol{\upvartheta}_m)^\mathsf{T} $ and applying the corresponding \emph{adjoint} quantum master equation to each of them; i.e.,
\begin{equation}\label{eq:closed-set}
    \frac{\D \boldsymbol{\upvartheta}_i}{\D t} = i[\pmb H_S, \boldsymbol{\upvartheta}_i] + \sum\nolimits_\alpha \mathcal{D}_\alpha^\dagger(\boldsymbol{\upvartheta}_i), 
\end{equation}
where the dissipators $ \mathcal{D}^\dagger_\alpha $ can take, e.g., the global $ \mathcal{G}^\dagger_\alpha $, local $ \mathcal{L}^\dagger_\alpha $, or Redfield $ \mathcal{R}^\dagger_\alpha $ form (cf. \eqref{eq:global_dissipator}, \eqref{eq:local_dissipator}, and \eqref{eq:Redfield}).

The aim is to pick observables $ \boldsymbol{\upvartheta}_j $ so that \eqref{eq:closed-set} becomes a \emph{closed} set of equations \cite{insinga2018quantum}. This can then be expressed in compact form as
\begin{equation}\label{eq:system-of-eqs}
    \frac{\D \langle\bsftheta\rangle}{\D t} = \mathsf{M}_\mathcal{D}\,\langle\bsftheta\rangle,
\end{equation}
were the resulting matrix of coefficients $  \mathsf{M}_\mathcal{D} \in \mathbb{C}^{m\times m} $ is generally non-Hermitian. At an EP, $  \mathsf{M}_\mathcal{D} $ stops being diagonalizable, which has a detectable impact on the dynamics and thermodynamics of the open system \cite{am2015exceptional,insinga2018quantum,minganti2019quantum}.

Specifically, in continuous-variable settings with linear $\pmb{H}_S$---as our example model---it is always possible to write a set of equations like \eqref{eq:system-of-eqs}. For instance, we can build a 14-element $ \bsftheta $ by grouping the four position and momentum operators 
\begin{equation}
    \boldsymbol{\mathsf{q}} \coloneqq (\pmb x_h , \pmb p_h , \pmb x_c , \pmb p_c)^\mathsf{T},
\end{equation}
together with the $ 10 $ distinct combinations
\begin{equation}
    \boldsymbol{\mathsf{C}}_{ij} \coloneqq \frac12\{\boldsymbol{\mathsf{q}}_i,\boldsymbol{\mathsf{q}}_j\}_+,
\end{equation}
e.g., $ \boldsymbol{\mathsf{C}}_{33} = \pmb x_c^2$, $ \boldsymbol{\mathsf{C}}_{12} = \boldsymbol{\mathsf{C}}_{21} = \frac12\{ \pmb x_h,\pmb p_h \}_+ $, or $ \boldsymbol{\mathsf{C}}_{32} = \boldsymbol{\mathsf{C}}_{23} = \pmb x_c\,\pmb p_h $. Note that other choices of $ \bsftheta $ are possible. Ordering the observables so that $ \pmb x_h $, $ \pmb p_h $, $ \pmb x_c $, and $ \pmb p_c $ are the first elements of $ \bsftheta $ results in the $14\times 14$ coefficient matrix $  \mathsf{M}_{\mathcal{D}} $
\begin{equation}\label{eq:big-matrix-coefficients}
 \mathsf{M}_\mathcal{D} =  \begin{pmatrix}
   \mathsf{M}_{\mathcal{D}, 1}  & \rvline & 0 \\
\hline
  0 & \rvline &   \mathsf{M}_{\mathcal{D},2}
\end{pmatrix}.
\end{equation}
That is, for the Hamiltonian $ \pmb H_S $, one finds that $  \mathsf{M}_\mathcal{D} $ is the direct sum of sub-matrices $  \mathsf{M}_{\mathcal{D},1}\in\mathbb{C}^{4\times 4} $ and $  \mathsf{M}_{\mathcal{D},2}\in\mathbb{C}^{10\times 10} $. Looking back at Eq.~\eqref{eq:system-of-eqs}, we thus see that the dynamics of the first-order moments $ \langle \boldsymbol{\mathsf{q}}_i \rangle $ decouples from that of the second-order moments $ \langle \boldsymbol{\mathsf{C}}_{ij} \rangle $. Note that the same block-diagonal structure is found for the local ($\mathcal{L} $), the global ($\mathcal{G} $), and Redfield ($\mathcal{R} $) equations. Since the exceptional points are properties of the dissipator \cite{minganti2019quantum}, it seems reasonable that their signatures appear both in $\mathsf{M}_{\mathcal{D},1}$ and $\mathsf{M}_{\mathcal{D},2}$. In Sec.~\ref{sec:results-and-discussion} below, we focus on the appearance of EPs in the sub-matrices $\mathsf{M}_{\mathcal{D},1}$ while the discussion about $\mathsf{M}_{\mathcal{D},2}$ has been deferred to Appendix~\ref{app:GME_and_LME_second_order}.


\section{Results and discussion}\label{sec:results-and-discussion}

\subsection{Local master equation}

\begin{figure*}
    \centering
    \includegraphics[width=\textwidth]{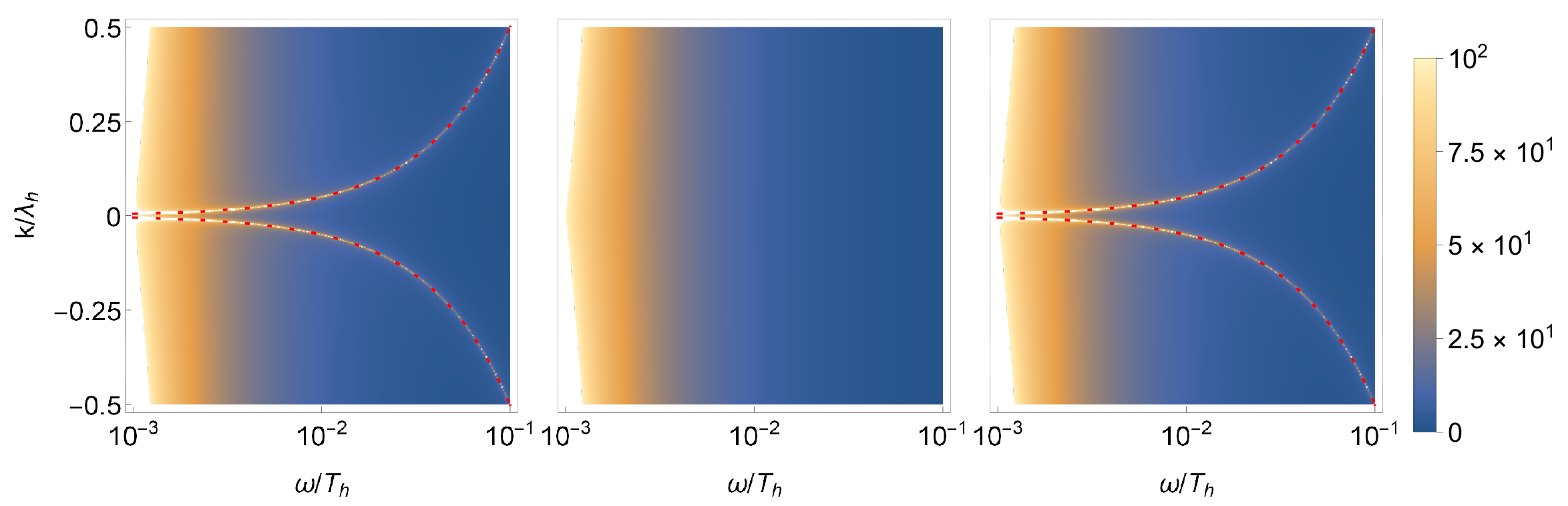}
    \caption{\textbf{EPs in the dynamics of first-order moments.} Condition number $ \upkappa $ (cf.~\eqref{eq:condition-number}) of the matrix of eigenvectors of $  \mathsf{M}_{\mathcal{L},1} $ \textbf{(left)}, $  \mathsf{M}_{\mathcal{G},1} $ \textbf{(centre)}, and $  \mathsf{M}_{\mathcal{R},1} $ \textbf{(right)}, as a function of the frequency of the resonators $ \omega_h = \omega_c = \omega $ and the strength of their capacitive coupling $ k $ (see Eqs.~\eqref{eq:coeffs-first-order-lme}, \eqref{eq:coeffs_gme-original-basis-matrix}, and \eqref{eq:coefficients-redfield-first-order-matrix}). These are the coefficient matrices of the equations of motion for the first-order moments $ \langle\pmb{\mathsf{q}}\rangle $ of the coupled-resonator system, according to the local, global, and Redfield master equation, respectively. A diverging condition number is the distinct signature of an exceptional point. The EPs predicted by the Redfield equation (right) are lost as a result of the secular approximation and thus, entirely missed by the global approach (middle). They are, however, captured by the local master equation (left). Note that the EPs appear exactly along the `exceptional lines' \eqref{eq:exceptional-line}, superimposed in dashed red. The parameters chosen are $ T_h = 10 $, $ T_c = 5 $, $ \lambda_h^2 = 10^{-8}$, $ \lambda_c^2 = 10^{-4}$, and $ \Lambda = 10^3$ ($ \hbar = k_B = 1 $).}\label{fig1}
\end{figure*}

In order to obtain the equations of motion for the first-order moments $ \langle\pmb{\mathsf{q}}\rangle $ of the system according to the LME, we need to know the Bohr frequencies involved, and the corresponding jump operators. From now on, we work in the regime of resonant oscillators, i.e. $ \omega_h = \omega_c = \omega $, which is a necessary condition for the appearance of the exceptional points in the example model considered, as shown in Appendix~\ref{app:resonant-asymmetric-condition}. 

The Bohr frequencies in Eq.~\eqref{eq:local_dissipator} thus become $ \{\omega'\}= \{\pm \omega\} $, and the local jump operators of \eqref{eq:local_dissipators_commutation} become
\begin{align}
    \pmb L_{\omega}^{(\alpha)} = \frac{\pmb a_\alpha}{\sqrt{2\omega}},\quad\pmb L_{-\omega}^{(\alpha)} = \pmb L_{\omega}^{(\alpha)\,\dagger},
\end{align}
where $ \pmb a_\alpha $ is an annihilation operator, so that $ \pmb x_\alpha = \pmb L_\omega^{(\alpha)} + \pmb L_{-\omega}^{(\alpha)} = (\pmb a_\alpha + \pmb a_\alpha^\dagger)/\sqrt{2\omega}$. Replacing these into the LME \eqref{eq:local_dissipator} gives
\begin{equation}\label{eq:first-order-lme}
    \frac{\D \langle\pmb{\mathsf{q}}\rangle}{\D t} =  \mathsf{M}_{\mathcal{L},1}\langle\pmb{\mathsf{q}}\rangle,
\end{equation}
with local coefficient matrix $  \mathsf{M}_{\mathcal{L},1} $
\begin{equation}\label{eq:coeffs-first-order-lme}
    \mathsf{M}_{\mathcal{L},1} = 
    \frac12\begin{pmatrix}
    \Delta_\omega^{(h)} & 2 & 0 & 0 \\
    -2\omega^2 & \Delta_\omega^{(h)} & - 2 k & 0 \\
    0 & 0 & \Delta_\omega^{(c)} & 2 \\
    - 2k & 0 & -2\omega^2 & \Delta_\omega^{(c)}
    \end{pmatrix},
\end{equation}
where we have introduced the notation
\begin{equation}\label{eq:deltas_def}
    \Delta_\omega^{(\alpha)} \coloneqq \frac{\gamma_{-\omega}^{(\alpha)}-\gamma_{\omega}^{(\alpha)}}{2\omega}.
\end{equation}

Since $  \mathsf{M}_{\mathcal{L},1} $ is simple enough, one can obtain an analytic expression for the exceptional points in parameter space. Looking at its eigenvalues, we see that degeneracy appears, provided that
\begin{multline}
    \label{eq:exceptional-line_raw}
    4 k^2 - \big(\Delta_\omega^{(h)}-\Delta_\omega^{(c)}\big)^2 \omega^2 = 0 \\
    \implies k = \pm\,\frac{\omega}{2}\,\vert \Delta_\omega^{(h)}-\Delta_\omega^{(c)} \vert.
\end{multline}
Further replacing the expressions of the decay rates $ \gamma_{\pm\omega}^{(\alpha)} $ into the coefficients $ \Delta_\alpha $ for our choice of spectral density (cf. Eqs.~\eqref{eq:ohmic-algebraic} and \eqref{eq:decay_rate_system}) results in the remarkably simple expression
\begin{equation}\label{eq:exceptional-line}
    k = \pm\,\frac{\omega}{2}\,\vert \lambda_h^2 - \lambda_c^2 \vert.
\end{equation}
Resorting now to the condition number of the eigenvector matrix of $  \mathsf{M}_{\mathcal{L},1} $, we confirm that whole family of exceptional points does lie along \eqref{eq:exceptional-line} (see leftmost panel Fig.~\ref{fig1}). That is, adjusting resonance frequency and internal couplings, it is always possible to tune the system into an EP. Interestingly, for exceptional points to appear in this system, dissipation must be asymmetric and the oscillators \emph{resonant} (see Appendix~\ref{app:resonant-asymmetric-condition}). Note as well that, at resonance, the LME cannot give rise to unphysical cold-to-hot heat currents \cite{levy2014local}. Moreover, in a recent experiment with coupled superconducting resonators, signatures of precisely the EPs in \eqref{eq:exceptional-line} have been indeed detected \cite{partanen2019exceptional}. 

\subsection{Global master equation}

In order to find the jump operators $ \pmb A_\omega^{(\alpha)} $ within the global dissipators $ \mathcal{G}_\alpha^\dagger $ (cf. \eqref{eq:global_dissipator} and Eqs.~\eqref{eq:jump_operators_properties}), we must rotate our system into its normal-mode quadratures $ \pmb{\mathsf{Q}} = (\pmb\eta_1, \pmb\Pi_1,\pmb\eta_2,\pmb\Pi_2)^\mathsf{T} $, so that
\begin{equation}
    \pmb H_S = \sum_{j=1}^2 \left(\frac{\pmb \Pi_j^2}{2} + \frac12 \Omega_j^2\pmb\eta_j^2\right).
\end{equation}
In resonance, the orthogonal transformation $ (\pmb{x}_h, \pmb{x}_c)^\mathsf{T} = \mathsf{P}\,(\pmb{\eta}_1, \pmb{\eta}_2)^\mathsf{T} $ between local and global modes has the form
\begin{equation}\label{eq:change-of-basis}
    \mathsf{P} = 
    \frac{1}{\sqrt{2}}\left(\begin{matrix}
    1 & 1 \\
    1 & -1 
    \end{matrix}\right),
\end{equation}
and the normal-mode frequencies are
\begin{equation}
    \Omega_{1,2} = \sqrt{\omega^2 \pm k}.
\end{equation}

We must decompose the system's coupling operators $ \pmb x_\alpha $ (cf. Eq.~\eqref{eq:h_int}) in eigenoperators of $ \pmb H_S $. Taking, for instance, $ \pmb x_h $, one can see that
\begin{equation}
    \pmb x_h = \frac{\pmb\eta_1 + \pmb\eta_2}{\sqrt{2}} = \frac{\pmb b_1}{2\sqrt{\Omega_1}} + \frac{\pmb b_2}{2\sqrt{\Omega_2}}  + \text{h.c.},
\end{equation}
where $ \pmb b_j $ is the annihilation operator associated with $ \pmb \eta_j $. Hence, 
\begin{align}\label{eq:global_jump_operators}
    \pmb A_{\Omega_j}^{(h)} &= \frac{\mathsf{P}_{1j}}{\sqrt{2\Omega_j}}\,\pmb b_j = \frac{\mathsf{P}_{1j}}{2}\left( \pmb \eta_j + \frac{\iu}{\Omega_j}\,\pmb \Pi_j \right) ,\\
    \pmb A_{\Omega_j}^{(c)} &= \frac{\mathsf{P}_{2j}}{\sqrt{2\Omega_j}}\,\pmb b_j = \frac{\mathsf{P}_{2j}}{2}\left( \pmb \eta_j + \frac{\iu}{\Omega_j}\,\pmb \Pi_j \right).
\end{align}
Also note that, unlike in the local approach, now there are two open decay channels into each bath, at frequencies $ \Omega_1 $ and $ \Omega_2 $, respectively. 

We can obtain the equations of motion for the normal-mode variables $ \langle\pmb{\mathsf{Q}}\rangle $ from \eqref{eq:global_dissipator};
\newpage
namely,
\begin{align}\label{eq:coeffs_gme}
    \frac{\D \langle\pmb{\mathsf{Q}}\rangle}{\D t} &=  \mathsf{M}'_{\mathcal{G},1}\langle\pmb{\mathsf{Q}}\rangle \nonumber, \\
     \mathsf{M}'_{\mathcal{G},1} &= \bigoplus_{j=1}^2
    \begin{pmatrix}
    \tilde{\Delta}_j/2 & 1 \\
    -\Omega_j^2 & \tilde{\Delta}_j/2
    \end{pmatrix},
\end{align}
where we have introduced the new coefficients
\begin{equation}\label{eq:deltas_gme}
    \tilde{\Delta}_j \coloneqq \frac12\sum\nolimits_{\alpha} \Delta_{\Omega_j}^{(\alpha)} .
\end{equation}
For completeness, we can rotate Eq.~\eqref{eq:coeffs_gme} back to the original coordinates $ \pmb{\mathsf{q}} $, which gives
\begin{widetext}
\begin{subequations}
\begin{align}
    \frac{\D \langle\pmb{\mathsf{q}}\rangle}{\D t} &=  \mathsf{M}_{\mathcal{G},1}\langle\pmb{\mathsf{q}}\rangle, \label{eq:coeffs_gme-original-basis-eqs}\\
    \mathsf{M}_{\mathcal{G},1} &= 
    \frac12\begin{pmatrix}
    \frac12(\tilde\Delta_1+\tilde\Delta_2) & 2 & \frac12(\tilde\Delta_1 - \tilde\Delta_2) & 0 \\
    -2\omega^2 & \frac12(\tilde\Delta_1 + \tilde\Delta_2) & -2k & \frac12 (\tilde\Delta_1-\tilde\Delta_2) \\
    \frac12 (\tilde\Delta_1-\tilde\Delta_2) & 0 & \frac12(\tilde\Delta_1+\tilde\Delta_2) & 2 \\
    -2k & \frac12 (\tilde\Delta_1-\tilde\Delta_2) & -2\omega^2 & \frac12(\tilde\Delta_1 + \tilde\Delta_2)\label{eq:coeffs_gme-original-basis-matrix}
    \end{pmatrix}.
\end{align}
\end{subequations}
\end{widetext}
The coefficient matrices $  \mathsf{M}'_{\mathcal{G},1} $ and $  \mathsf{M}_{\mathcal{G},1} $ have the same condition number $ \upkappa $, since they are connected via an orthogonal transformation. As a result, they also have the same EPs, since the norms involved in the calculation of $ \upkappa $ remain unaffected (see Eq.~\eqref{eq:condition-number}).

As seen in the middle panel of Fig.~\ref{fig1}, the `exceptional lines' of diverging condition number in the frequency--coupling space \emph{disappear completely}, according to the global master equation.

One may question the \emph{validity} of the GME for such parameters. Namely, in Fig.~\ref{fig1} we set $ k \sim \lambda^2_\alpha $ whereas, to be on the safe side when it comes to the secular approximation, we should ensure instead that $ k\gg \max_\alpha\lambda^2_\alpha $ \cite{gonzalez2017testing}. However, as we show in Sec.~\ref{sec:heat-currents}, the GME does lead to the correct steady-state properties at all plotted points save for the fringe $ \vert k \vert \lessapprox 0.1\,\lambda_h^2 $. Hence the disappearance of the EPs cannot be simply attributed to the global approach breaking down.

Thinking of the local coefficient matrix $  \mathsf{M}_{\mathcal{L},1} $ as the lowest order  $ \mathsf{M}_{\mathcal{G},1}^{(0)}$ of a perturbative expansion in $ k $ of the dissipative contributions to $  \mathsf{M}_{\mathcal{G},1}$, i.e.
\begin{equation*}
    \mathsf{M}_{\mathcal{G},1} =  \mathsf{M}_{\mathcal{G},1}^{(0)} + k\, \mathsf{M}_{\mathcal{G},1}^{(1)} + \cdots ,
\end{equation*}
it would even be tempting to disregard the EP singularities predicted by the LME as mathematical artifacts, and trust instead in the \emph{a priori} more physical GME. However, as advanced in Sec.~\ref{sec:localvsglobal}, this interpretation is \emph{not} valid here. To see why, we only need to calculate $  \mathsf{M}_{\mathcal{G},1}^{(0)}$ from Eq.~\eqref{eq:coeffs_gme-original-basis-matrix} and show that it differs from $  \mathsf{M}_{\mathcal{L},1} $ in Eq.~\eqref{eq:coeffs-first-order-lme}. Namely, we must set $ k = 0 $ in all terms arising from the global dissipators $ \mathcal{G}_\alpha^\dagger$ \eqref{eq:global_dissipator}, while keeping those from the commutator part of Eq.~\eqref{eq:GKSL_general} intact \cite{trushechkin2016perturbative}. This is achieved by replacing all coefficients $ \tilde{\Delta}_j $ in $ \mathsf{M}_{\mathcal{G},1}$ by $ (\Delta_\omega^{(h)} + \Delta_\omega^{(c)})/2$, i.e.,
\begin{widetext}
\begin{equation}\label{eq:local-not-derived-from-global}
     \mathsf{M}_{\mathcal{G},1}^{(0)} = 
    \frac12\begin{pmatrix}
    \frac12(\Delta_\omega^{(h)} + \Delta_\omega^{(c)}) & 2 & 0 & 0 \\
    -2\omega^2 & \frac12(\Delta_\omega^{(h)} + \Delta_\omega^{(c)}) & - 2k & 0 \\
    0 & 0 & \frac12(\Delta_\omega^{(h)} + \Delta_\omega^{(c)}) & 2 \\
    - 2k & 0 & -2\omega^2 & \frac12(\Delta_\omega^{(h)} + \Delta_\omega^{(c)})
    \end{pmatrix} \neq  \mathsf{M}_{\mathcal{L},1}.
\end{equation}
\end{widetext}
That is, even in the limit $ k = 0 $, each heat bath continues to act globally on \emph{both} resonators, rather than on the resonator directly coupled to it. This is a side-effect of the secular approximation, which introduces a `heat leak' channel. And that is why the GME predicts steady-state heat flows in the resonant case even at vanishing coupling, as noted in Refs.~\cite{gonzalez2017testing,Hofer2017}. 

Eq.~\eqref{eq:local-not-derived-from-global} confronts us with the fact that the LME is \emph{not}, in general, a limiting case of the GME. But then, \emph{what is it?} What we are after is a \emph{microscopic justification} of the local equation \eqref{eq:first-order-lme} capable of explaining why it succeeds in capturing the EPs detected experimentally in \cite{partanen2019exceptional}, while the global equation \eqref{eq:coeffs_gme-original-basis-eqs} fails. 

\subsection{Redfield equation}\label{sec:redfield-equation}

We now resort to the Redfield equation, as introduced in Eqs. \eqref{eq:redfield_main_text} and \eqref{eq:redfield_main_text_2}, to shed light on the nature of the LME. As outlined in Sec.~\ref{sec:global_master_equation}, this equation is the last step in the derivation of the GME, before forcing the GKSL form by means of the (full) secular approximation. Consequently, the Redfield equation is \emph{always} more accurate than the GME. In the context of quantum thermodynamics, however, the main shortcoming of the Redfield equation is its lack of GKSL structure. Without it, it may fail to generate a completely positive dynamics \cite{suarez1992memory,gaspard1999slippage}. Nonetheless, if used with caution, the Redfield equation can still yield thermodynamically sound results \cite{jeske2015bloch,strasberg2016nonequilibrium,hartman2020embracing}.

To be more precise, \eqref{eq:redfield_main_text} is a \emph{partial Redfield equation} \cite{jeske2015bloch,cresser2017coarse,gonzalez2017testing,cattaneo2019local}. That is, a simplified version of the equation in which its most rapidly oscillating terms---but \emph{not all} oscillating terms---are averaged out under a coarse graining of time. Below, we also discard the \emph{Lamb shifts}, defined in Appendix~\ref{app:clean_derivation_master_equation}. The resulting system is
\begin{subequations}
    \begin{align}\label{eq:coeffs-eqs-redfield}
        \frac{\D \langle \pmb{\mathsf{q}} \rangle}{\D t} &=  \mathsf{M}_{\mathcal{R},1}\,\langle \pmb{\mathsf{q}} \rangle,\\
        \mathsf{M}_{\mathcal{R},1} &= \frac12 \label{eq:coefficients-redfield-first-order-matrix}
        \begin{pmatrix}
            [\mathsf{M}_{\mathcal{R},1}]_{11} & 2 & [ \mathsf{M}_{\mathcal{R},1}]_{13} & 0\\
            -2\omega^2 & \bar{\Delta}_h & -2k & \delta_h \\
            [ \mathsf{M}_{\mathcal{R},1}]_{31} & 0 & [ \mathsf{M}_{\mathcal{R},1}]_{33} & 2 \\
            -2\omega^2 & \delta_c & -2\omega^2 & \bar{\Delta}_c
        \end{pmatrix},
    \end{align}
where $ \delta_\alpha \coloneqq \frac12(\Delta_1^{(\alpha)}-\Delta_2^{(\alpha)}) $, ~ $ \bar{\Delta}_\alpha \coloneqq \frac12(\Delta_1^{(\alpha)}+\Delta_2^{(\alpha)}) $, and
    \begin{widetext}
    \begin{align}\label{eq:matrix-elements-Redfield}
        [ \mathsf{M}_{\mathcal{R},1}]_{11} &= \frac{\Omega_2-\Omega_1}{4}\left(\frac{\Delta_1^{(c)}}{\Omega_2} - \frac{\Delta_2^{(c)}}{\Omega_1} \right) + \frac{\Omega_2+\Omega_1}{4}\left(\frac{\Delta_1^{(h)}}{\Omega_2} + \frac{\Delta_2^{(h)}}{\Omega_1} \right) \nonumber\\
        [ \mathsf{M}_{\mathcal{R},1}]_{13} &= \frac{\Omega_2-\Omega_1}{4}\left(\frac{\Delta_1^{(c)}}{\Omega_2} + \frac{\Delta_2^{(c)}}{\Omega_1} \right) + \frac{\Omega_2+\Omega_1}{4}\left(\frac{\Delta_1^{(h)}}{\Omega_2} - \frac{\Delta_2^{(h)}}{\Omega_1} \right) \nonumber\\
        [ \mathsf{M}_{\mathcal{R},1}]_{31} &=  \frac{\Omega_2+\Omega_1}{4}\left(\frac{\Delta_1^{(c)}}{\Omega_2} - \frac{\Delta_2^{(c)}}{\Omega_1} \right)\nonumber + \frac{\Omega_2-\Omega_1}{4}\left(\frac{\Delta_1^{(h)}}{\Omega_2} - \frac{\Delta_2^{(h)}}{\Omega_1} \right)\\
        [ \mathsf{M}_{\mathcal{R},1}]_{33} &=  \frac{\Omega_2+\Omega_1}{4}\left(\frac{\Delta_1^{(c)}}{\Omega_2} + \frac{\Delta_2^{(c)}}{\Omega_1} \right) + \frac{\Omega_2-\Omega_1}{4}\left(\frac{\Delta_1^{(h)}}{\Omega_2} + \frac{\Delta_2^{(h)}}{\Omega_1} \right).
    \end{align}
    \end{widetext}
\end{subequations}

\begin{figure*}[t!]
    \centering
    \includegraphics[width=.45\linewidth]{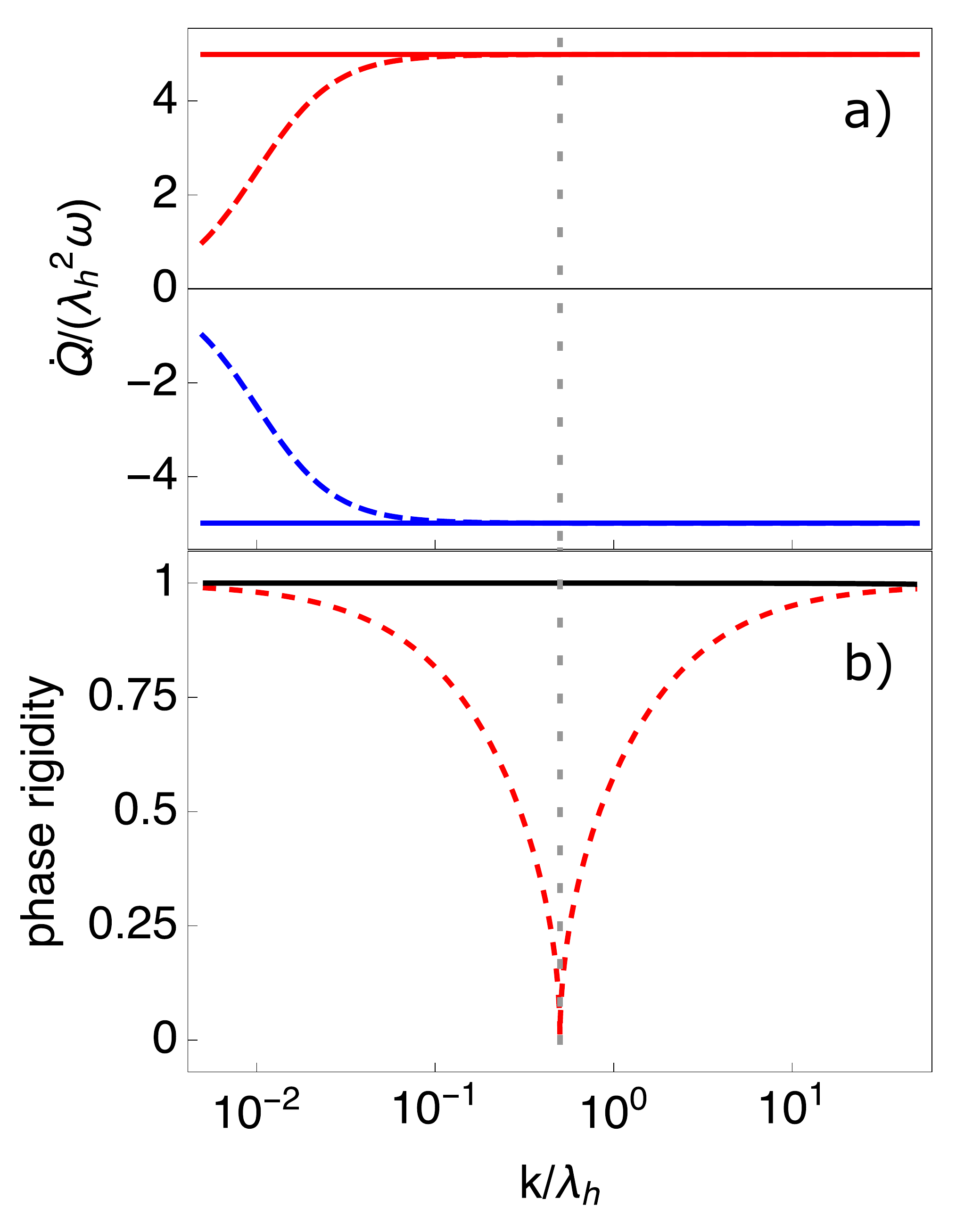}
    \hfill%
    \includegraphics[width=.45\linewidth]{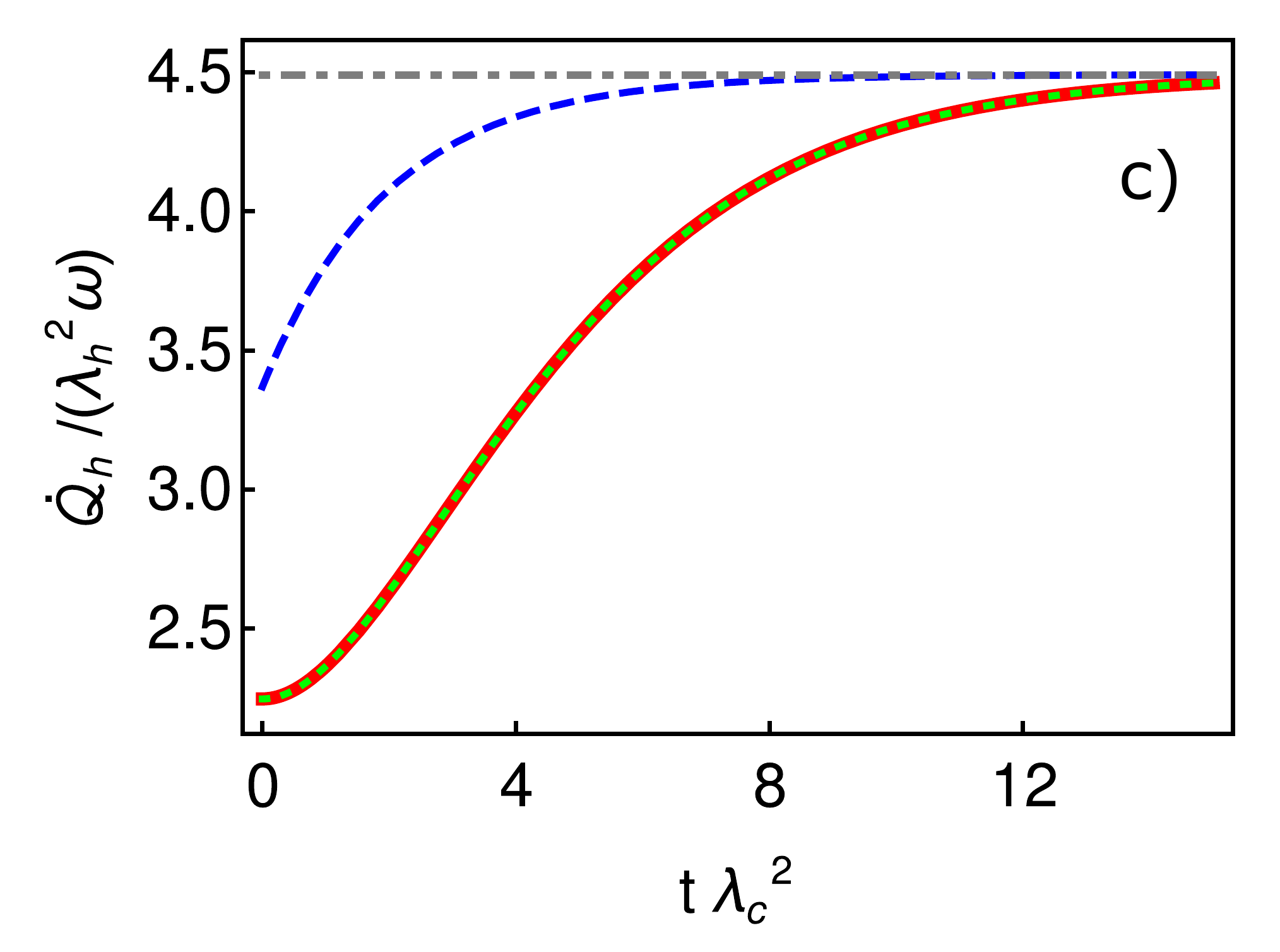}
    \caption{\textbf{Heat currents at an exceptional point.} \textbf{a)} Steady-state heat currents from the hot bath (red) and the cold bath (blue) as a function of $ k $, according to the GME (solid) and the LME (dashed). \textbf{b)} Phase rigidity calculated according to the LME (dashed red) for the two coalescing eigenvectors, and according to the GME (solid black), for which no eigenvectors coalesce\footnote{Note that the phase rigidity seems to converge to unity (solid black) away from the EP. Recall from Sec.~\ref{sec:EPs-definition-witnesses} that $ \upphi_i = 1 $ only for Hermitian systems. In fact, evaluating the behaviour of $\upphi_i$ in the range of parameters shown in the plot, we see that the phase rigidity remains $ \upphi_i < 1 $, albeit by a very small amount ($\sim 10^{-5}$)}. The position of the EP according to \eqref{eq:exceptional-line} is indicated by the dotted grey line across 
    \textbf{a)} and \textbf{b)}. \textbf{c)} Transient of the hot heat current $ \dot{\pazocal{Q}}_h(t) $ according to the LME (solid red), the GME (dashed blue), and the Redfield equation (dotted green). The steady-state value is indicated by the dot-dashed grey line. The global approach thus deviates from the true heat-flow dynamics, set by the Redfield equation. On the contrary, the local approach remains accurate throughout. Parameters are as in Fig.~\ref{fig1}, except for $ \omega = 1 $. As initial state, we take the tensor product of the thermal state of each oscillator at its local bath's temperature.}
    \label{fig3}
\end{figure*}

In spite of the cumbersome expressions, one can see that the zeroth order term $  \mathsf{M}_{\mathcal{R},1}^{(0)} $ of the $ k $-expansion of the dissipative part of $  \mathsf{M}_{\mathcal{R},1} $,
\begin{equation}
     \mathsf{M}_{\mathcal{R},1} =  \mathsf{M}_{\mathcal{R},1}^{(0)} + k\, \mathsf{M}_{\mathcal{R},1}^{(1)} + k^2\, \mathsf{M}_{\mathcal{R},1}^{(2)}\cdots,
\end{equation}
is identical to $  \mathsf{M}_{\mathcal{L},1} $ in Eq.~\eqref{eq:coeffs-first-order-lme} (or, equivalently, $ \mathcal{R}_\alpha^{\dagger\,(0)} = \mathcal{L}_\alpha^\dagger $, see Appendix~\ref{app:connecting-LME-with-GME-out-of-resonance}). One only needs to set $ k = 0 $ in all dissipative contributions of Eq.~\eqref{eq:redfield_main_text}, which is equivalent to setting $ k = 0 $ in the $ \bar{\Delta}_\alpha $ and $ \delta_\alpha $ terms, as well as in the four matrix elements written out in Eqs.~\eqref{eq:matrix-elements-Redfield}.

We have thus shown that, in resonance, \emph{the LME is the low-$ k $ limit of the Redfield equation; not of the GME}. In fact, it is easy to see that $ \mathcal{R}_\alpha^{\dagger\,(0)} = \mathcal{L}_\alpha^\dagger $ holds as well out of resonance for \emph{any} multipartite model, as long as the coupling to the heat baths is mediated by `frequency filters'. More precisely, these are multipartite systems in which the coupling $ \pmb H_{SB} $ is of the form, e.g., $\pmb{a}_\alpha\otimes\pmb B_\alpha + \text{h.c.} $ or $ \ket{\epsilon_i}\bra{\epsilon_j}\otimes\pmb B_\alpha + \text{h.c.} $, where $ \ket{\epsilon_i} $ are eigenstates of the local Hamiltonian $ \pmb H_\alpha^{(\text{loc})}$ (see Appendix~\ref{app:connecting-LME-with-GME-out-of-resonance} for details). In such settings, the LME emerges \emph{directly} from the Redfield equation. This is why $  \mathsf{M}_{\mathcal{R},1} $ and $  \mathsf{M}_{\mathcal{L},1} $ share the same pattern of EPs in parameter space (cf. Fig.~\ref{fig1}).

Crucially, unlike the Redfield approach, the LME is guaranteed to generate a completely positive dissipative dynamics \cite{mozgunov2020completely}. In addition, this explains the unlikely success of the LME over the GME at low $ k $, when the secular approximation breaks down \cite{gonzalez2017testing,Hofer2017}. In this new light, we see that the LME simply \emph{bypasses} the secular approximation. This is one of the main results of this paper.

Furthermore, note from Eq.~\eqref{eq:change-of-basis} that the eigenstates of $ \pmb H_S $ do \emph{not} depend on $ k $. It then becomes clear why the local approach remains accurate even at larger couplings in resonance \cite{Hofer2017}: A small-$k$ approximation of the dissipators $ \mathcal{R}_\alpha^\dagger $ (or $ \mathcal{G}_\alpha^\dagger $) is valid over a wider range of couplings if the expansion affects only the Bohr frequencies $ \omega $, but not the jump operators $ \pmb A_\omega^{(\alpha)} $. Conversely, out of resonance, the LME loses validity at non-zero couplings, since the eigenstates of $ \pmb H_S $ are then explicitly dependent on $ k $ (see Appendix~\ref{app:connecting-LME-with-GME-out-of-resonance}). We note as well that EPs can be studied exploiting the stochastic unraveling of GKSL master equations to define an effective non-Hermitian Hamiltonian, as shown in \cite{minganti2019quantum}.

The failure of the GME at capturing the correct dynamics is precisely due to the fact that the secular approximation---required by the GME---`washes away' relevant \emph{dynamical} features. We have just illustrated this with the disappearance of the EPs. Next, we show that, even when the LME and GME do agree in their steady-state predictions, the local approach can generate more accurate heat-flow dynamics.

\subsection{Heat currents}\label{sec:heat-currents}

The \emph{instantaneous} rate of heat flow into the system from each of the baths can be obtained by generalising Eq.~\eqref{eq:heat-current-definition} to
\begin{equation}\label{eq:instantaneous-heat}
    \dot{\pazocal{Q}}_{\alpha,\mathcal{D}}(t) \coloneqq \langle \mathcal{D}_\alpha^\dagger(\pmb H_S)\rangle_t,
\end{equation}
where the transient heat currents $ \dot{\pazocal{Q}}_\alpha(t) $ generally \emph{do not} sum up to zero, nor obey the Clausius-like inequality \eqref{eq:clausius-ineq}; they only do so at very long times, $ t\rightarrow\infty $. Directly applying this definition using the local, global, and Redfield dissipators gives
\begin{align}\label{eq:heat-currents}
    \dot{\pazocal{Q}}_{\alpha,\mathcal{L}} &=  \frac{\Delta_\omega^{(\alpha)}}{2}\big(\langle \pmb p_\alpha^2 \rangle +\omega^2\langle \pmb x_\alpha^2 \rangle + k\langle\pmb x_h\pmb x_c \rangle\big) \nonumber + \frac{\Sigma_\omega^{(\alpha)}}{2}, \nonumber\\
    \dot{\pazocal{Q}}_{\alpha,\mathcal{G}} &=  \frac14\sum\nolimits_j\big[\Delta^{(\alpha)}_{\Omega_j}\big(\langle\pmb\Pi_j^2\rangle + \Omega_j^2 \langle\pmb\eta_j^2\rangle \big) \nonumber + \Sigma^{(\alpha)}_{\Omega_j}\big], \nonumber\\
    \dot{\pazocal{Q}}_{\alpha,\mathcal{R}} &= \dot{\pazocal{Q}}_{\alpha,\mathcal{G}} \nonumber\\  &+ \frac{\Delta_1^{(\alpha)} + \Delta_2^{(\alpha)}}{4}\big(\Omega_1\Omega_2 \langle \pmb \eta_1\pmb\eta_2 \rangle + \langle\pmb\Pi_1\pmb\Pi_2 \rangle\big),
\end{align}
where $ \Sigma_\omega^{(\alpha)} \coloneqq \frac{1}{2\omega}(\gamma_{-\omega}^{(\alpha)} + \gamma_\omega^{(\alpha)}) $, and $ \langle \pmb\eta_i\pmb\eta_j \rangle $ and $ \langle \pmb\Pi_i\pmb\Pi_j \rangle $ are second-order moments. Therefore, in order to evaluate Eqs.~\eqref{eq:heat-currents} we must set up and solve the corresponding linear system of 10 equations with coefficient matrices $  \mathsf{M}_{2,\mathcal{D}} $ from Eq.~\eqref{eq:big-matrix-coefficients}. This is a rather tedious but, otherwise, straightforward process. As it turns out, the $ 10\times 10 $ matrices $  \mathsf{M}_{\mathcal{L},2} $ and $  \mathsf{M}_{\mathcal{R},2} $ have the same exceptional points than their $ 4\times 4 $ first-order counterparts $  \mathsf{M}_{\mathcal{L},1} $ and $  \mathsf{M}_{\mathcal{R},1} $.
We defer details to Appendix~\ref{app:GME_and_LME_second_order}. 

In Fig.~\ref{fig3} we tune the parameters to be at an exceptional point of the system according to Eq.~\eqref{eq:exceptional-line}, and plot both steady-state and transient heat currents. For the chosen parameters, LME and GME agree in their steady-state predictions to a very good approximation, see Fig.~\hyperref[fig3]{3a)}. However, local and global dynamics do differ significantly at finite time. In contrast, the Redfield equation agrees with the LME  at all times, see Fig.~\hyperref[fig3]{3c)}. This suggests that, within its error bars, \emph{the local approach can be superior to the global one when studying the thermodynamics of multipartite systems with weak internal couplings}.


\section{Conclusions}\label{sec:conclusions}

We have analysed one of the central problems in quantum thermodynamics. Namely, the modelling of heat flows across open systems with quantum master equations. We have shown that the two most common approaches---the local and the global master equations---can make very different predictions. Firstly, our results illustrate that the local approach succeeds at capturing dynamical features, in the form of exceptional points, that escape the global master equation. Secondly, we have found that, when considering degenerate multipartite open quantum systems with weak internal coupling, the LME also yields much more accurate heat-flow dynamics than the GME, even when both agree in the steady-state.

Furthermore, we have shown that the LME follows directly from the more accurate Redfield equation, and is generally \emph{not} a weak-coupling limit of the GME. This is always the case for any multipartite weakly-interacting open quantum system that couples to the environment(s) via single-frequency contacts, such as a qubit or a harmonic oscillator. Therefore, for such systems, the LME emerges as an accurate and computationally efficient alternative to the Redfield equation. It proves to be superior to the GME and, in contrast to the Redfield equation, it does guarantee positivity. 

These results have profound consequences for quantum thermodynamics. Namely, modelling heat flow in a quantum thermal machine with the local approach instead of the global one, could make a sizeable difference in the predicted heat transfer in, e.g., any thermalising stroke of a finite-time thermodynamic cycle. This could result in a radically different assessment of both performance and power output. However, it is important to remember that the local approach has a limited range of validity. Specifically, it is unsuitable for open systems with \emph{strong} internal couplings (i.e., large $k$), and it can lead to unphysical results at odds with thermodynamics. Finding an accurate and scalable master equation for such scenarios still remains an open challenge. 

\section*{Source code}
The code used to produce the figures is available upon reasonable request to \href{mailto:s.scali@exeter.ac.uk}{s.scali@exeter.ac.uk}.

\acknowledgements{We gratefully acknowledge J. Cresser, S. Horsley, G. de Chiara, P. Strasberg, and G. Haack for helpful comments. SS is supported by a DTP grant from EPSRC (EP/R513210/1). JA gratefully acknowledges support from EPSRC (EP/R045577/1) and the Royal Society.}

\bibliography{refs}

\clearpage

\onecolumn
\appendix
\numberwithin{equation}{section}

\section{Derivation of the GME}\label{app:clean_derivation_master_equation}

Here, we derive a second-order master equation under the Markov and secular approximations. Essentially, we follow Ref.~\cite{gaspard1999slippage}, but simplify the derivation. Let the Hamiltonian of our system be
\begin{equation}
    \pmb H = \pmb H_S + \pmb H_B + \lambda \pmb H_{SB},
\end{equation}
where $ \pmb H_{SB} $ stands for the dissipative interactions between system (S) and bath (B). Note that we are transferring the magnitude of these interactions into $ \lambda $, which means that the rescaled $ \pmb H_{SB} $ is now $\pazocal{O}(1)$, unlike in the main text (cf. Eq.~\eqref{eq:h_int}).   

Our starting point will be the Liouville--von Neumann equation in the interaction picture with respect to $ \pmb H_0 \coloneqq \pmb H_S + \pmb H_B $. This is
\begin{equation}\label{eq:Liouville-von_Neumann}
    \frac{\D\pmb{\tilde\rho}}{\D t} = -i\,\lambda\,[\pmb{\tilde{H}}_{SB}(t),\pmb{\tilde\rho}(t)] \coloneqq \lambda\,\tilde{\pazocal{L}}(t)\,\pmb{\tilde\rho}(t),
\end{equation}
where $ \pmb{\tilde{O}}(t) \coloneqq e^{i\pmb H_0 t}\,\pmb O\,e^{-i\pmb H_0 t} $. Here, $ \pmb{\tilde{\rho}}(t) $ is the full system--environment state. Since we are only interested in the system's marginal $ \pmb{\tilde{\varrho}} \coloneqq \tr_B \pmb{\tilde{\rho}}(t)$, we trace out the bath, i.e., 
\begin{equation}
    \frac{\D\pmb{\tilde{\varrho}}}{\D t} = \lambda\tr_B\tilde{\pazocal{L}}(t)\,\pmb{\tilde{\rho}}(t)
\end{equation}
and integrate formally. This gives us
\begin{equation}\label{eq:intermediate1}
    \pmb{\tilde{\varrho}}(t) = \pmb\varrho(0) + \lambda\int_0^t \D s\,\tr_B\tilde{\pazocal{L}}(s)\pmb{\tilde{\rho}}(s), 
\end{equation}
and iterating,
\begin{equation}\label{eq:intermediate2}
    \pmb{\tilde{\varrho}}(t) = \pmb \varrho(0) + \lambda\int_0^t \D s \tr_B\tilde{\pazocal{L}}(s)\pmb\rho(0)
    + \lambda^2 \int_0^t \D s \int_0^s \D s' \tr_B\tilde{\pazocal{L}}(s)\tilde{\pazocal{L}}(s') \pmb{\tilde{\rho}}(s').
\end{equation}
Replacing the state $ \pmb{\tilde{\rho}}(s') $ in Eq.~\eqref{eq:intermediate2} by the expression in \eqref{eq:intermediate1} once again, we see that
\begin{equation}\label{eq:intermediate3}
    \pmb{\tilde{\varrho}}(t) = \pmb \varrho(0) + \lambda \int_0^t \D s \,\tr_B\tilde{\pazocal{L}}(s)\pmb\rho(0) 
    + \lambda^2 \int_0^t \D s \int_0^s \D s' \,\tr_B\tilde{\pazocal{L}}(s)\tilde{\pazocal{L}}(s') \pmb\rho(0) \\+ \pazocal{O}(\lambda^3).
\end{equation}
Assuming that $ \lambda $ is small enough so that all $ \pazocal{O}(\lambda^3) $ terms or smaller can be neglected will be our first key approximation. This will lead to a \emph{second-order quantum master equation} after taking the time derivative; namely, 
\begin{equation}\label{eq:weak-coupling-master-equation}
    \frac{\D \pmb{\tilde{\varrho}}}{\D t} = \lambda \,\tr_B\tilde{\pazocal{L}}(t)\pmb{\rho}(0) + \lambda^2 \int_0^t \D s \,\tr_B\tilde{\pazocal{L}}(t)\tilde{\pazocal{L}}(s) \pmb\rho(0).
\end{equation}

Notice, however, that this master equation time-non-local and, still, of little practical use. To overcome this problem, we make two additional assumptions: First, we require the \emph{initial} state to include no correlations between system's and bath's degrees of freedom. Namely, $\pmb \rho(0) = \pmb \varrho_S(0)\otimes\pmb\varrho_B$. Secondly, we impose 
\begin{equation}\label{eq:homogeneous}
    \tr_B\tilde{\pazocal{L}}(s)\pmb{\varrho}(0)\otimes\pmb\varrho_B = 0.
\end{equation}
For instance, if the system--bath interaction is of the simple form $ \pmb H_{SB} = \pmb S\otimes\pmb B $ (cf. Eq.~\eqref{eq:h_int}), Eq.~\eqref{eq:homogeneous} translates into $ \langle\pmb B\rangle = 0 $ when averaged on the initial state of the environment. 

Inserting \eqref{eq:homogeneous} into \eqref{eq:intermediate3}, we see that $ \pmb\varrho(0) = \pmb{\tilde{\varrho}}(t) + \pazocal{O}(\lambda^2) $, which allows us to write our equation in the much more convenient time-local form
\begin{equation}
    \frac{\D \pmb{\tilde{\varrho}}}{\D t} = \lambda^2 \int_0^t \D s \,\tr_B\tilde{\pazocal{L}}(t)\tilde{\pazocal{L}}(s) \tilde{\pmb{\varrho}}(t)\otimes\pmb\varrho_B
\end{equation}
while still remaining accurate within $\pazocal{O}(\lambda^3)$. It is customary to perform a change of variables in the integral, re-expressing it as
\begin{equation}\label{eq:time-local-second-order}
    \frac{\D \pmb{\tilde{\varrho}}}{\D t} = \lambda^2 \int_0^t \D s \,\tr_B\tilde{\pazocal{L}}(t)\tilde{\pazocal{L}}(t-s) \pmb{\tilde{\varrho}}(t)\otimes\pmb\varrho_B.
\end{equation}

We note that objects like
\begin{equation}
    \tr_B\tilde{\pazocal{L}}(t)\tilde{\pazocal{L}}(t-s) \pmb{\tilde{\varrho}}(t)\otimes\pmb\varrho_B
\end{equation}
enclose two-time correlation functions of bath operators, i.e., $ \tr_B\pmb{\tilde{B}}(t)\pmb{\tilde{B}}(t-s)\pmb\varrho_B $. In many situations of practical interest, these decay extremely fast---much faster than any relevant timescale in the problem. Hence, we may replace the upper integration limit in \eqref{eq:time-local-second-order} by infinity without substantial error
\begin{equation}\label{eq:second-order-homogeneous-time-local-Markov}
    \frac{\D \pmb{\tilde{\varrho}}}{\D t} = \lambda^2 \int_0^\infty \D s \,\tr_B\tilde{\pazocal{L}}(t)\tilde{\pazocal{L}}(t-s) \pmb{\tilde{\varrho}}(t)\otimes\pmb\varrho_B.    
\end{equation}

This is often referred-to as the \emph{Markov approximation} and the resulting equation, as (Markovian) \emph{Redfield master equation}. However, the use of the term `Markovian' can be problematic since the dynamics generated by this equation is, in general, not even \emph{positive} \cite{gaspard1999slippage}. This means that the corresponding dynamical map might not be \emph{divisible}, and lack of divisibility is often associated with `\emph{non}-Markovianity' \cite{rivas2010entanglement}. 

Before the final step in our derivation, we must transcribe the shorthand $ \tr_B\tilde{\pazocal{L}}(t)\tilde{\pazocal{L}}(t-s) \pmb{\tilde{\varrho}}(t)\otimes\pmb\varrho_B $, which gives
\begin{equation*}
    \tilde{\pazocal{L}}(t)\tilde{\pazocal{L}}(t-s) \pmb{\tilde{\varrho}}(t)\otimes\pmb\varrho_B =
    -\pmb{\tilde{H}}_{SB}(t)\pmb{\tilde{H}}_{SB}(t-s)\pmb{\tilde{\varrho}}(t)\otimes\pmb\varrho_B + \pmb{\tilde{H}}_{SB}(t-s)\pmb{\tilde{\varrho}}(t)\otimes\pmb\varrho_B\pmb{\tilde{H}}_{SB}(t) + \text{h.c.} 
\end{equation*}
We may always write the system--bath interaction term as $ \pmb{H}_{SB} = \sum_\alpha\pmb{S}_\alpha\otimes\pmb B_\alpha $. Here, we take $ \pmb H_{SB} = \pmb S\otimes\pmb B$ for simplicity, and hence,
\begin{equation*}
    \tilde{\pazocal{L}}(t)\tilde{\pazocal{L}}(t-s) \pmb{\tilde{\varrho}}(t)\otimes\pmb\varrho_B =
    -\pmb{\tilde{S}}(t)\pmb{\tilde{S}}(t-s)\pmb{\tilde{\varrho}}(t)\,\tr_B\pmb{\tilde{B}}(t)\pmb{\tilde{B}}(t-s)\pmb\varrho_B
    + \pmb{\tilde{S}}(t-s)\pmb{\tilde{\varrho}}(t)\pmb{\tilde{S}}(t)\,\tr_B\pmb{\tilde{B}}(t-s)\pmb\varrho_B\pmb{\tilde{B}}(t) + \text{h.c.}
\end{equation*}

The explicit form of the interaction-picture system operator $ \pmb{\tilde{S}}(t) $ can be found easily by exploiting the decomposition from Eqs.~\eqref{eq:jump_operators_properties} in the main text. Namely, given the properties of the jump operators $ \pmb A_\omega $ it is easy to see that $ \pmb{\tilde{S}}(t) = \sum\nolimits_\omega \pmb A_\omega\,e^{-i\omega t} $. Putting together all the pieces  
\begin{multline}\label{eq:Redfield-almost-final}
    \frac{\D\pmb{\tilde{\varrho}}}{\D t} =
    \sum\nolimits_{\omega,\omega'}\left(-\Gamma(\omega)\,\pmb{A}_{\omega'}\pmb{A}_{\omega}e^{-i(\omega + \omega') t}\pmb{\tilde{\varrho}}(t) \right.
    \left.+ \Gamma(\omega)\,\pmb{A}_\omega\pmb{\tilde{\varrho}}(t)\pmb{A}_{\omega'}e^{-i(\omega+\omega') t} + \text{h.c.}\right) \\= \sum\nolimits_{\omega,\omega'}\left(-\Gamma(\omega)\,\pmb{A}_{\omega'}^\dagger\pmb{A}_{\omega}e^{i(\omega' - \omega) t}\pmb{\tilde{\varrho}}(t) \right.
    \left.+ \Gamma(\omega)\,\pmb{A}_\omega\pmb{\tilde{\varrho}}(t)\pmb{A}_{\omega'}^\dagger e^{i(\omega'-\omega') t} + \text{h.c.}\right),
\end{multline}
where
\begin{equation}
    \Gamma(\omega) \coloneqq \lambda^2\int\nolimits_0^\infty \D s\, e^{i\omega s} \tr_B \pmb{\tilde{B}}(t)\pmb{\tilde{B}}(t-s),
\end{equation}
and the corresponding real and imaginary parts are $ \Gamma(\omega) = \frac12\gamma(\omega) + i S(\omega)$. The imaginary part $ S(\omega) $ is typically ignored. It introduces two effects---a displacement of the energy levels of $ \pmb H_S $ through a \emph{Lamb shift} term  $ \pmb H_{L} = \sum\nolimits_\omega S(\omega) \pmb A_\omega^\dagger \pmb A_\omega $ ($ [\pmb H_S,\pmb H_L] = 0 $); but also, non-trivial dissipative terms. These are, however, typically very small.  

Many terms in Eq.~\eqref{eq:Redfield} can be dropped, since they are fast-oscillating and average out to zero over the time-scale defined by the dynamics of $ \pmb \varrho(t) $ \cite{breuer2002theory}. Namely, we can drop all terms for which $ \omega $ and $ \omega' $ have the same sign ($ \omega\times\omega' > 0 $), since these oscillate as $ e^{\pm \vert \omega + \omega' \vert t} $. Eq.~\eqref{eq:Redfield-almost-final} is often called `partial Redfield equation' \cite{cresser2017coarse,gonzalez2017testing,cattaneo2019local}

It is now time to abandon the interaction picture undoing the corresponding unitary transformation. This gives us
\begin{equation}\label{eq:Redfield}
    \frac{\D\pmb{\varrho}}{\D t} = -i[\pmb H_S,\pmb{\varrho}] + \frac12\sum_{\omega\times\omega' < 0}\gamma(\omega)\left(\pmb{A}_\omega\pmb{\varrho}(t)\pmb{A}_{\omega'}^\dagger-\pmb{A}_{\omega'}^\dagger\pmb{A}_{\omega}\pmb{\varrho}(t)\right) + \text{h.c.}
\end{equation}

One final step is necessary to bring Eq.~\eqref{eq:Redfield} into GKSL form---the \emph{secular approximation}. This consists in removing all terms in which $ \omega \neq \omega' $ from the double sum in \eqref{eq:Redfield}. The rationale for this---seemingly arbitrary---simplification often involves again a time averaging which now `kills' \emph{all} the remaining oscillating terms in Eq.~\eqref{eq:Redfield-almost-final} \cite{breuer2002theory}. From all approximations involved in the process, this is certainly the most problematic and difficult to justify. Nonetheless, it does hold in many situations of practical interest \cite{breuer2002theory}. This way, we finally arrive at the celebrated global GKSL master equation
\begin{equation}\label{eq:global}
    \frac{\D\pmb{\varrho}}{\D t} = -i[\pmb H_S,\pmb{\varrho}]
    + \sum_{\omega}\gamma(\omega)\left(\pmb{A}_\omega\pmb{\varrho}(t)\pmb{A}_{\omega}^\dagger-\frac12\{\pmb{A}_{\omega}^\dagger\pmb{A}_{\omega}\pmb{\varrho}(t)\}_+\right).
\end{equation}

\section{Dynamics of second-order moments}
\label{app:GME_and_LME_second_order}
We now study the dynamics of the second-order moments showing the matrices of the dynamics generated by LME (Eq.~\eqref{eq:2nd_order_lme}), GME (Eq.~\eqref{eq:2nd_order_gme}) and Redfield equation (Eq.~\eqref{eq:2nd_order_redfield}).

We first consider the local approach. Using Eq.~\eqref{eq:LME}, we obtain a set of 10 coupled first-order differential equations defining the dynamics of the covariances $\pmb{\mathsf{C}}_{ij}$ of the system, as defined in Sec.~\ref{sec:EPs-open-systems}. These equations can be expressed as
\begin{equation}
    \frac{\D \langle\pmb{\tilde{\mathsf{q}}}\rangle}{\D t} =  \mathsf{M}_{\mathcal{L},2}\langle\pmb{\tilde{\mathsf{q}}}\rangle + \mathsf{c}_{\mathcal{L},2},
\end{equation}
where the ordered vector of the covariances is
\begin{align}
    \boldsymbol{\tilde{\mathsf{q}}} \coloneqq \left( \pmb x_h^2 , \pmb p_h^2 , \frac12\{ \pmb x_h,\pmb p_h \}_+ , \pmb x_c^2 , \pmb p_c^2, \frac12\{ \pmb x_c,\pmb p_c \}_+ ,\pmb x_h \pmb p_c , \pmb x_c \pmb p_h , \pmb x_h \pmb x_c , \pmb p_h \pmb p_c \right)^\mathsf{T},
\end{align}
and the constant vector $\mathsf{c}_{\mathcal{L},2}$ is
\begin{equation}
    \mathsf{c}_{\mathcal{L},2} = \left( \frac{\Sigma_\omega^{(h)}}{2\omega}, \frac{\omega\Sigma_\omega^{(h)}}{2}, 0, \frac{\Sigma_\omega^{(c)}}{2\omega}, \frac{\omega\Sigma_\omega^{(c)}}{2}, 0, 0, 0, 0, 0 \right)^\mathsf{T}.
\end{equation}
The coefficients $\Sigma_\omega^{(\alpha)}$s are defined as
\begin{equation}\label{eq:sigmas_def}
    \Sigma_\omega^{(\alpha)} \coloneqq \frac{\gamma_{-\omega}^{(\alpha)}+\gamma_{\omega}^{(\alpha)}}{2\omega}.
\end{equation}
Therefore, in the local approach, the matrix of the dynamics takes the form
\begin{equation}\label{eq:2nd_order_lme}
     \mathsf{M}_{\mathcal{L},2} = 
    \begin{pmatrix}
    \Delta_\omega^{(h)} & 0 & 2 & 0 & 0 & 0 & 0 & 0 & 0 & 0\\
    0 & \Delta_\omega^{(h)} & -2\omega^2 & 0 & 0 & 0 & 0 & k & 0 & 0 \\
    -\omega^2 & 1 & \Delta_\omega^{(h)} & 0 & 0 & 0 & 0 & 0 & k/2 & 0 \\
    0 & 0 & 0 & \Delta_\omega^{(c)} & 0 & 2 & 0 & 0 & 0 & 0 \\
    0 & 0 & 0 & 0 & \Delta_\omega^{(c)} & -2\omega^2 & k & 0 & 0 & 0 \\
    0 & 0 & 0 & -\omega^2 & 1 & \Delta_\omega^{(c)} & 0 & 0 & k/2 & 0 \\
    k/2 & 0 & 0 & 0 & 0 & 0 & \frac{\Delta_\omega^{(h)}+\Delta_\omega^{(c)}}{2} & 0 & -\omega^2 & 1 \\
    0 & 0 & 0 & k/2 & 0 & 0 & 0 & \frac{\Delta_\omega^{(h)}+\Delta_\omega^{(c)}}{2} & -\omega^2 & 1 \\
    0 & 0 & 0 & 0 & 0 & 0 & 1 & 1 & \frac{\Delta_\omega^{(h)}+\Delta_\omega^{(c)}}{2} & 0 \\
    0 & 0 & k/2 & 0 & 0 & k/2 & -\omega^2 & -\omega^2 & 0 & \frac{\Delta_\omega^{(h)}+\Delta_\omega^{(c)}}{2}
    \end{pmatrix},
\end{equation}
where the coefficients $\Delta_\omega^{(\alpha)}$ were already introduced in Eq.~\eqref{eq:deltas_def}. Studying (numerically) the condition number of the eigenvectors matrix of $ \mathsf{M}_{\mathcal{L},2}$, we obtain the exact same line of singularities emerging from the condition in Eq.~\eqref{eq:exceptional-line}.

Let us consider next the case of the global master equation. Since the condition number is independent of the basis, in the following we report the dynamics expressed in terms of the normal-mode quadratures of the system. In the case of the global master equation, we obtain 
\begin{equation}
    \frac{\D \langle\pmb{\tilde{\mathsf{Q}}}\rangle}{\D t} =  \mathsf{M}'_{\mathcal{G},2}\langle\pmb{\tilde{\mathsf{Q}}}\rangle + \mathsf{c}'_{\mathcal{G},2}.
\end{equation}
In this case, the ordered vector of the covariances is
\begin{align}
    \boldsymbol{\tilde{\mathsf{Q}}} \coloneqq \left( \pmb \eta_1^2 , \pmb \Pi_1^2 , \frac12\{ \pmb \eta_1,\pmb \Pi_1 \}_+ , \pmb \eta_2^2 , \pmb \Pi_2^2, \frac12\{ \pmb \eta_2,\pmb \Pi_2 \}_+ ,\pmb \eta_1 \pmb \Pi_2 , \pmb \eta_2 \pmb \Pi_1 , \pmb \eta_1 \pmb \eta_2 , \pmb \Pi_1 \pmb \Pi_2 \right)^\mathsf{T},
\end{align}
and the constant vector is given by
\begin{equation}
    \mathsf{c}'_{\mathcal{G},2} = \left( \frac{\tilde\Sigma_1}{2\Omega_1}, \frac{\Omega_1\tilde\Sigma_1}{2}, 0, \frac{\tilde\Sigma_2}{2\Omega_2}, \frac{\Omega_2\tilde\Sigma_2}{2}, 0, 0, 0, 0, 0 \right)^\mathsf{T},
\end{equation}
while the matrix of the dynamics is
\begin{equation}\label{eq:2nd_order_gme}
     \mathsf{M}'_{\mathcal{G},2} = 
    \begin{pmatrix}
    \tilde\Delta_1 & 0 & 2 & 0 & 0 & 0 & 0 & 0 & 0 & 0\\
    0 & \tilde\Delta_1 & -2\Omega_1^2 & 0 & 0 & 0 & 0 & 0 & 0 & 0\\
    -2\Omega_1^2 & 1 & \tilde\Delta_1 & 0 & 0 & 0 & 0 & 0 & 0 & 0\\
    0 & 0 & 0 & \tilde\Delta_2 & 0 & 2 & 0 & 0 & 0 & 0\\
    0 & 0 & 0 & 0 & \tilde\Delta_2 & -2\Omega_2^2 & 0 & 0 & 0 & 0\\
    0 & 0 & 0 & -\Omega_2^2 & 1 & \tilde\Delta_2 & 0 & 0 & 0 & 0\\
    0 & 0 & 0 & 0 & 0 & 0 & \frac{\tilde\Delta_1+\tilde\Delta_2}{2} & 0 & -\Omega_2^2 & 1\\
    0 & 0 & 0 & 0 & 0 & 0 & 0 & \frac{\tilde\Delta_1+\tilde\Delta_2}{2} & -\Omega_1^2 & 1\\
    0 & 0 & 0 & 0 & 0 & 0 & 1 & 1 & \frac{\tilde\Delta_1+\tilde\Delta_2}{2} & 0\\
    0 & 0 & 0 & 0 & 0 & 0 & -\Omega_1^2 & -\Omega_2^2 & 0 & \frac{\tilde\Delta_1+\tilde\Delta_2}{2}
    \end{pmatrix}.
\end{equation}
The $\tilde{\Delta}_j$ have been introduced in Eq.~\eqref{eq:deltas_gme} and, analogously, we define here the coefficients
\begin{equation}\label{eq:sigmas_gme}
    \tilde{\Sigma}_j \coloneqq \frac12\sum\nolimits_{\alpha} \Sigma_{\Omega_j}^{(\alpha)}.
\end{equation}

In this case, the evaluation of the condition number reveals no exceptional points, just as in the case of the first-order moments. Hence, the discrepancy between the local and global master equation persists at the level of the second-order moments.

Now, we present the evolution of the covariances in the case of the Redfield equation. Again, in normal-mode variables, the dynamics is expressed as
\begin{equation}
    \frac{\D \langle\pmb{\tilde{\mathsf{Q}}}\rangle}{\D t} =  \mathsf{M}'_{\mathcal{R},2}\langle\pmb{\tilde{\mathsf{Q}}}\rangle + \mathsf{c}'_{\mathcal{R},2}.
\end{equation}
The matrix of the dynamics takes the form
\begin{equation}\label{eq:2nd_order_redfield}
     \mathsf{M}'_{\mathcal{R},2} = 
    \begin{pmatrix}
    \tilde\Delta_1 & 0 & 2 & 0 & 0 & 0 & 0 & 0 & \frac{\Omega_2}{\Omega1}\tilde\Delta_2' & 0\\
    0 & \tilde\Delta_1 & -2\Omega_1^2 & 0 & 0 & 0 & 0 & 0 & 0 & \tilde\Delta_2'\\
    -2\Omega_1^2 & 1 & \tilde\Delta_1 & 0 & 0 & 0 & \frac{\tilde\Delta_2'}{2} & \frac{\Omega_2}{2\Omega_1}\tilde\Delta_2' & 0 & 0\\
    0 & 0 & 0 & \tilde\Delta_2 & 0 & 2 & 0 & 0 & \frac{\Omega_1}{\Omega_2}\tilde\Delta_1' & 0\\
    0 & 0 & 0 & 0 & \tilde\Delta_2 & -2\Omega_2^2 & 0 & 0 & 0 & \tilde\Delta_1'\\
    0 & 0 & 0 & -\Omega_2^2 & 1 & \tilde\Delta_2 & \frac{\tilde\Delta_1'}{2} & 0 & 0 & 0\\
    0 & 0 & \frac{\tilde\Delta_1'}{2} & 0 & 0 & \frac{\Omega_2}{2\Omega_1}\tilde\Delta_2' & \frac{\tilde\Delta_1+\tilde\Delta_2}{2} & 0 & -\Omega_2^2 & 1\\
    0 & 0 & \frac{\Omega_1}{2\Omega_2}\tilde\Delta_1' & 0 & 0 & \frac{\tilde\Delta_2'}{2} & 0 & \frac{\tilde\Delta_1+\tilde\Delta_2}{2} & -\Omega_1^2 & 1\\
    \frac{\Omega_1}{2\Omega_2}\tilde\Delta_1' & 0 & 0 & \frac{\Omega_2}{2\Omega_1}\tilde\Delta_2' & 0 & 0 & 1 & 1 & \frac{\tilde\Delta_1+\tilde\Delta_2}{2} & 0\\
    0 & \frac{\tilde\Delta_1'}{2} & 0 & 0 & \frac{\tilde\Delta_2'}{2} & 0 & -\Omega_1^2 & -\Omega_2^2 & 0 & \frac{\tilde\Delta_1+\tilde\Delta_2}{2}
    \end{pmatrix},
\end{equation}
while the constant vector is
\begin{equation}
    \mathsf{c}'_{\mathcal{R},2} = \left( \frac{\tilde\Sigma_1}{2\Omega_1}, \frac{\Omega_1\tilde\Sigma_1}{2}, 0, \frac{\tilde\Sigma_2}{2\Omega_2}, \frac{\Omega_2\tilde\Sigma_2}{2}, 0, 0, 0, \frac{\tilde\Sigma_1'}{4\Omega_2} + \frac{\tilde\Sigma_2'}{4\Omega_1}, \frac{\Omega_1\tilde\Sigma_1'}{4} + \frac{\Omega_2\tilde\Sigma_2'}{4} \right)^\mathsf{T}.
\end{equation}
In these expressions, we have defined the new coefficients
\begin{subequations}
\begin{align}\label{eq:deltas_redfield}
    \tilde{\Delta}_j' &\coloneqq \frac12 \left(\Delta_{\Omega_j}^{(h)} - \Delta_{\Omega_j}^{(c)} \right) \\
    \tilde{\Sigma}_j' &\coloneqq \frac12 \left(\Sigma_{\Omega_j}^{(h)} - \Sigma_{\Omega_j}^{(c)} \right).\label{eq:sigmas_redfield}
\end{align}
\end{subequations}
Calculating the condition number of the corresponding matrix of eigenvectors, one can readily confirm, once again, the presence of the exceptional points along the exact same `exceptional lines' \eqref{eq:exceptional-line}. 

It is worth noting that, since the system considered is Gaussian, these singularities will be present, according to the LME and the Redfield equation, in any $n^\text{th}$-order moment. This is so, because any higher-order correlation functions of a Gaussian system can be cast as a combination of first- and second-order moments, and these simultaneously display exceptional points. Conversely, no moments of any order will ever pick up exceptional points according to the global description.

\section{Conditions for the appearance of exceptional points}
\label{app:resonant-asymmetric-condition}
As mentioned in the main text, for the exceptional points to appear, dissipation must be asymmetric and the oscillators must be resonant. Here, we show the validity of this statement.

Consider the non-resonant case, where the resonators have bare frequencies $\omega_h$ and $\omega_c$. The matrix of the dynamics of the first order moments is
\begin{equation}
    \tilde{\mathsf{M}}_{\mathcal{L},1} = 
    \frac12\begin{pmatrix}
    \Delta_{\omega_h}^{(h)} & 2 & 0 & 0 \\
    -2\omega_h^2 & \Delta_{\omega_h}^{(h)} & - 2 k & 0 \\
    0 & 0 & \Delta_{\omega_c}^{(c)} & 2 \\
    - 2k & 0 & -2\omega_c^2 & \Delta_{\omega_c}^{(c)}
    \end{pmatrix},
\end{equation}
where
\begin{equation}
    \Delta_{\omega_i}^{(\alpha)} \coloneqq \frac{\gamma_{-\omega_i}^{(\alpha)}-\gamma_{\omega_i}^{(\alpha)}}{2\omega_i}.
\end{equation}
To identify the exceptional points we look for square root singularities in the expressions of the eigenvalues. Two of the eigenvalues of the latter matrix are
\begin{align}
    \lambda_{1,2} = \frac{1}{4}\Big[ \Delta_{\omega_h}^{(h)} + \Delta_{\omega_c}^{(c)} &+2\sqrt{-(\omega_h + \omega_c)^2} + \\ \nonumber
    &\pm\sqrt{(\Delta_{\omega_h}^{(h)} - \Delta_{\omega_c}^{(c)})^2 - 4(\omega_h-\omega_c)^2 - \frac{4k^2}{\omega_h \omega_c} - 4(\Delta_{\omega_h}^{(h)}-\Delta_{\omega_c}^{(c)})\frac{\omega_h^2-\omega_c^2}{\sqrt{-(\omega_h+\omega_c)^2}}} \Big]
\end{align}
which coalesce when the latter square root vanishes. Therefore, we find the condition of exceptional points for the inter-resonators coupling $k$,
\begin{equation}
    k = \pm \frac{1}{2} \sqrt{\omega_h \omega_c} \sqrt{(\Delta_{\omega_h}^{(h)} - \Delta_{\omega_c}^{(c)})^2 -4 (\omega_h - \omega_c)\left[ (\omega_h - \omega_c) - i (\Delta_{\omega_h}^{(h)} - \Delta_{\omega_c}^{(c)}) \right]} .
\end{equation}
Since the inter-resonators coupling $k$ is assumed to be real, we have the only acceptable solution when the system is resonant, $\omega_h=\omega_c=\omega$ and dissipation asymmetric, $\Delta_h\neq\Delta_c$. In this way, we recover the expression in Eq.~\eqref{eq:exceptional-line_raw}.

\section{Local, global, and Redfield dissipators out of resonance}\label{app:connecting-LME-with-GME-out-of-resonance}
We now show that the relationship between the Redfield and the local approach pertains even in the non-resonant case. First of all, note that, out of resonance, the transformation matrix $ \mathsf{P} $ that defines the rotation into normal modes generalises to \cite{gonzalez2017testing}
\begin{equation}\label{eq:change-of-basis-out-of-resonance}
    \mathsf{P} = 
    \left(\begin{matrix}
    \sin{\zeta} & \cos{\zeta} \\
    \cos{\zeta} & -\sin{\zeta}
    \end{matrix}\right),
\end{equation}
where $\displaystyle\zeta=\arccos{\sqrt{\frac{\delta+\sqrt{4k^2+\delta^2}}{2\sqrt{4k^2+\delta^2}}}}$ and $\delta=\omega_h^2-\omega_c^2$. The eigenfrequencies are
\begin{equation}\label{eq:eigenvalues-out-of-resonance}
    \Omega_{1,2}^2=\frac12\left( \omega_h^2+\omega_c^2\pm\sqrt{4k^2+\delta^2}\right).
\end{equation}
Namely, in the non-resonant case \emph{both} the eigenvectors and the Bohr frequencies of $ \pmb H_S $ depend explicitly on the internal coupling strength $ k $.

Moving now to the Redfield dissipators \eqref{eq:Redfield} we can write it out in adjoint form as
\begin{align}\label{eq:redfield_dissipator_our_system}
    \mathcal{R}_h^\dagger(\pmb O) + \mathcal{R}_c^\dagger(\pmb O) = \sum_{i,j = 1}^2 \Big[
    &\frac{\gamma_{\Omega_i}^{(h)}}{2}\left( \pmb{A}_{-\Omega_j}^{(h)}\pmb O \pmb{A}_{\Omega_i}^{(h)} + \pmb{A}_{-\Omega_i}^{(h)}\pmb O \pmb{A}_{\Omega_j}^{(h)} - \pmb O \pmb{A}_{-\Omega_j}^{(h)}\pmb{A}_{\Omega_i}^{(h)} - \pmb{A}_{-\Omega_i}^{(h)}\pmb{A}_{\Omega_j}^{(h)}\pmb O \right)\\\notag
    +&\frac{\gamma_{-\Omega_i}^{(h)}}{2}\left( \pmb{A}_{\Omega_j}^{(h)}\pmb O \pmb{A}_{-\Omega_i}^{(h)} + \pmb{A}_{\Omega_i}^{(h)}\pmb O \pmb{A}_{-\Omega_j}^{(h)} - \pmb O \pmb{A}_{\Omega_j}^{(h)}\pmb{A}_{-\Omega_i}^{(h)} - \pmb{A}_{\Omega_i}^{(h)}\pmb{A}_{-\Omega_j}^{(h)}\pmb O \right)\\\notag
    +&\frac{\gamma_{\Omega_i}^{(c)}}{2}\left( \pmb{A}_{-\Omega_j}^{(c)}\pmb O \pmb{A}_{\Omega_i}^{(c)} + \pmb{A}_{-\Omega_i}^{(c)}\pmb O \pmb{A}_{\Omega_j}^{(c)} - \pmb O \pmb{A}_{-\Omega_j}^{(c)}\pmb{A}_{\Omega_i}^{(c)} - \pmb{A}_{-\Omega_i}^{(c)}\pmb{A}_{\Omega_j}^{(c)}\pmb O \right)\\\notag
    +&\frac{\gamma_{-\Omega_i}^{(c)}}{2}\left( \pmb{A}_{\Omega_j}^{(c)}\pmb O \pmb{A}_{-\Omega_i}^{(c)} + \pmb{A}_{\Omega_i}^{(c)}\pmb O \pmb{A}_{-\Omega_j}^{(c)} - \pmb O \pmb{A}_{\Omega_j}^{(c)}\pmb{A}_{-\Omega_i}^{(c)} - \pmb{A}_{\Omega_i}^{(c)}\pmb{A}_{-\Omega_j}^{(c)}\pmb O \right)\Big]
\end{align}
To find the zeroth order term $ \mathcal{R}^{\dagger\,(0)} $, we must simply set $ k = 0 $ in $ \mathsf{P} $ and the normal-mode frequencies. We thus find that the normal modes rotate back to the local coordinates $ \pmb x_\alpha $ and $ \Omega_\alpha $ collapse into the bare frequencies $ \omega_\alpha $. Importantly, as a result, frequency $ \Omega_1^{(0)} = \omega_h $ will only appear in the hot dissipator $ \mathcal{R}_h
^\dagger$ and $ \Omega_2^{(0)} = \omega_c $ will be only linked to $ \mathcal{R}^\dagger_c $. Therefore, the double sums in Eq.~\eqref{eq:redfield_dissipator_our_system} directly transform into the local expression \eqref{eq:local_dissipator} which in the specific case of the system studied is
\begin{align}\label{eq:local_dissipator_our_system}
    \mathcal{L}_h^\dagger(\pmb O) + \mathcal{L}_c^\dagger(\pmb O) = \;
    &\gamma_{\omega_h}^{(h)}\left( \pmb L_{-\omega_h}^{(h)}\pmb O \pmb L_{\omega_h}^{(h)} - \frac12\{ \pmb L^{(h)}_{-\omega_h} \pmb L^{(h)}_{\omega_h},\pmb O \}_+ \right) \\\notag
    +&\gamma_{-\omega_h}^{(h)}\left( \pmb L_{\omega_h}^{(h)}\pmb O \pmb L_{-\omega_h}^{(h)} - \frac12\{ \pmb L^{(h)}_{\omega_h} \pmb L^{(h)}_{-\omega_h},\pmb O \}_+ \right) \\\notag
    +&\gamma_{\omega_c}^{(c)}\left( \pmb L_{-\omega_c}^{(c)}\pmb O \pmb L_{\omega_c}^{(c)} - \frac12\{ \pmb L^{(c)}_{-\omega_c} \pmb L^{(c)}_{\omega_c},\pmb O \}_+ \right) \\\notag
    +&\gamma_{-\omega_c}^{(c)}\left( \pmb L_{\omega_c}^{(c)}\pmb O \pmb L_{-\omega_c}^{(c)} - \frac12\{ \pmb L^{(c)}_{\omega_c} \pmb L^{(c)}_{-\omega_c},\pmb O \}_+ \right) .
\end{align}
In the low-$k$ limit, this will always be true---regardless of $ \pmb H_S $---provided that the system couples to each bath by a single transition at some specific frequency.

In the same way, it is easy to show that the relationship between the global and the local dissipators, proposed in \cite{trushechkin2016perturbative}, remains valid in the out-of-resonance case. The adjoint global dissipator takes the form
\begin{align}\label{eq:global_dissipator_our_system}
    \mathcal{G}_h^\dagger(\pmb O) + \mathcal{G}_c^\dagger(\pmb O) = \sum_{i = 1}^2 \Big[
    &\gamma_{\Omega_i}^{(h)}\left( \pmb{A}_{-\Omega_i}^{(h)}\pmb O \pmb{A}_{\Omega_i}^{(h)} -\frac12 \{ \pmb{A}_{-\Omega_i}^{(h)}\pmb{A}_{\Omega_i}^{(h)}, \pmb O \}_+ \right)\\\notag
    +&\gamma_{-\Omega_i}^{(h)}\left( \pmb{A}_{\Omega_i}^{(h)}\pmb O \pmb{A}_{-\Omega_i}^{(h)} -\frac12 \{ \pmb{A}_{\Omega_i}^{(h)}\pmb{A}_{-\Omega_i}^{(h)}, \pmb O \}_+ \right)\\\notag
    +&\gamma_{\Omega_i}^{(c)}\left( \pmb{A}_{-\Omega_i}^{(c)}\pmb O \pmb{A}_{\Omega_i}^{(c)} -\frac12 \{ \pmb{A}_{-\Omega_i}^{(c)}\pmb{A}_{\Omega_i}^{(c)}, \pmb O \}_+ \right)\\\notag
    +&\gamma_{-\Omega_i}^{(c)}\left( \pmb{A}_{\Omega_i}^{(c)}\pmb O \pmb{A}_{-\Omega_i}^{(c)} -\frac12 \{ \pmb{A}_{\Omega_i}^{(c)}\pmb{A}_{-\Omega_i}^{(c)}, \pmb O \}_+ \right) \Big] .
\end{align}
Again, to find the zeroth order term of the $k$-expansion of the dissipator, we set $k=0$ in the matrix $\mathsf{P}$. Therefore, the global jump operators in Eq.~\eqref{eq:global_jump_operators} are cast to $ \pmb A_{\Omega_1}^{(h)} = \pmb L_{\omega_h}^{(h)}, \pmb A_{\Omega_2}^{(h)} = 0, \pmb A_{\Omega_1}^{(c)} = 0, \pmb A_{\Omega_2}^{(c)} = \pmb L_{\omega_c}^{(c)}$ which makes the global dissipator converge to the local expression.

\end{document}